\documentclass[useAMS,usenatbib,,usegraphicx]{mn2e}

\title[Extinction of extragalactic background light]{How the extinction of extragalactic background light affects surface photometry of galaxies, groups and clusters}
\author[Zackrisson et al.]{E. Zackrisson, G. Micheva, G. \"Ostlin\\
Oskar Klein Centre for Cosmoparticle Physics, Department of Astronomy, Stockholm University, 10691 Stockholm, Sweden\\
}

\begin{document}

\date{Accepted ... Received ...; in original form ...}

\pagerange{\pageref{firstpage}--\pageref{lastpage}} \pubyear{2009}

\maketitle

\label{firstpage}

\begin{abstract}
The faint regions of galaxies, groups and clusters hold important clues about how these objects formed, and surface photometry at optical and near-infrared wavelengths represents a powerful tool for studying such structures. Here, we identify a hitherto unrecognized problem with this technique, related to how the night sky flux is typically measured and subtracted from astronomical images. While most of the sky flux comes from regions between the observer and the target object, a small fraction –- the extragalactic background light (EBL) -– comes from behind. We argue that since this part of the sky flux can be subject to extinction by dust present in the galaxy/group/cluster studied, standard reduction procedures may lead to a systematic oversubtraction of the EBL. Even very small amounts of extinction can lead to spurious features in radial surface surface brightness profiles and colour maps of extended objects. We assess the likely impact of this effect on a number of topics in extragalactic astronomy where very deep surface photometry is currently attempted, including studies of stellar halos, starburst host galaxies, disc truncations and diffuse intragroup/intracluster light. We argue that EBL extinction may provide at least a partial explanation for the anomalously red colours reported for the halos of disc galaxies and the hosts of local starburst galaxies. EBL extinction effects also mimic truncations in discs with unusually high dust opacities, but are unlikely to be the cause of such features in general. Failure to account for EBL extinction can also give rise to a non-negligible underestimate of intragroup and intracluster light at the faintest surface brightness levels currently probed. Finally, we discuss how EBL extinction effects may be exploited to provide an independent constraint on the surface brightness of the EBL, using a combination of surface photometry and direct star counts.
\end{abstract}


\section{Introduction}
The low surface brightness regions of galaxies, galaxy groups and clusters hold important clues about how such systems formed. The faint stellar halos of galaxies are predicted to contain signatures of past mergers in the hierarchical build-up of galaxies \citep*[e.g.][]{Bullock & Johnston,Abadi et al.}, and intragroup and intracluster light may help us understand stripping and merger processes on larger scales \citep*[e.g.][]{Napolitano et al.,Murante et al.,Sommer-Larsen,Purcell et al.}. The outermost regions of galactic discs may also allow us to constrain the properties of star formation thresholds in the interstellar medium  \citep{Elmegreen & Parravano,Schaye}. 

At the current time, there are two complementary techniques for studying the faint regions of extended objects: direct star counts and surface photometry. By resolving bright, individual stars, it is possible to trace the outskirts of galaxies to remarkable distances from their centres, currently equivalent to surface brightness levels of $\mu_V\sim 35$ mag arcsec$^{-2}$ \citep[e.g.][]{Ibata et al.}. However, this technique is only applicable for nearby systems. Moreover, it is currently only the brightest stars -- with a very limited span of initial masses -- that are probed, whereas more diffuse flux components produced by the myriad of faint stars below the detection threshold and light emitted by the interstellar medium are not directly measured. Surface photometry offers a more concise census of the surface brightness, but is subject to a host of systematic problems which prevents regions as faint as those probed by star counts to be studied. 

The two chief problems which limit the depth of optical/near-IR surface photometry are instrumental scattering 
\citep{Michard,Sirianni et al.,de Jong} and the challenge of subtracting the night sky flux with sufficient accuracy \citep*[e.g.][]{Melnick et al.,Zheng et al.,Mihos et al.}. Because of these obstacles, optical surface photometry rarely probes regions faintward of the $\mu\approx 28$ mag arcsec$^{-2}$ isophote, although a few attempts have been made to push the limits into the $\mu\approx 30$--32 mag arcsec$^{-2}$ range \citep*[e.g.][]{Barton & Thompson,Zibetti et al. a, Zibetti et al. b}. Despite these limitations, surface photometry is a technique that we will have to live with for a long time to come.

Here, we identify a previously unrecognized problem with surface photometry at the faint limit, related to how the sky flux is typically estimated and subtracted from astronomical images. While most of the sky flux (in the form of airglow, zodiacal light, light from the stars and interstellar medium in the Milky Way) comes from regions between the observer and the object studied, a small fraction –- the extragalactic background light (EBL) -– comes from behind. The EBL at optical and infrared wavelengths is believed to be the product of direct and reprocessed starlight emitted over the entire star formation history of the Universe, and hence stems from objects at vastly different redshifts. Provided that the existing EBL measurements are correct, most of this light appears to be diffuse (i.e. unresolved with all existing instruments). Unlike the other components of the night sky flux, the EBL can be subject to extinction by dust present in the low surface brightness regions of the target object, thereby invalidating an implicit assumption in all current surface photometry measurements, namely that the sky flux and the flux from the target object are unrelated. 

In Section 2, we explain how this effect is likely to manifest itself in the surface brightness profiles of extended, low-redshift objects, arguing that spurious features are expected to turn up at surface brightness levels similar to or fainter than that of the EBL itself. Estimates of the surface brightness of the EBL in units of mag arcsec$^{-2}$ are presented in section 3 for Johnson-Cousins $UBVRIJHK$ and Sloan Digital Sky Survey (SDSS) $ugriz$ filters. In sections 4 through 7, we describe the relevance of EBL extinction for studies of disc truncations, halos of disc galaxies, host galaxies of local starbursts and intragroup/cluster light, respectively. In section 8, we outline how EBL extinction effects may be exploited to provide an independent measurement of the surface brightness of the EBL, using a combination of surface photometry and direct star counts. A number of remaining uncertainties in our analysis are discussed in section 9. Section 10 summarizes our findings.

\section{How extinction of extragalactic background light affects surface photometry} 
\begin{figure}
\includegraphics[width=84mm]{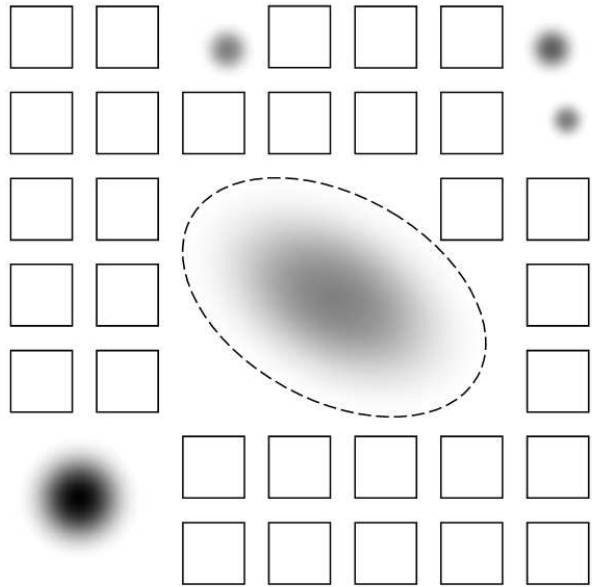}
\caption{Schematic illustration of how the surface brightness of the night sky is typically estimated. The sky level is measured in regions (boxes) well away from the target galaxy (central object) and other resolved sources in the frame. A surface or constant is then fitted to these measurements and subtracted from the frame. The problem with this procedure is that the sky level relevant for the target isophote (dashed ellipse) is likely to be slightly lower than in the regions where it was estimated. This happens because of EBL extinction by dust associated with the galaxy studied.}
\label{fig1}
\end{figure}
The surface brightness of the night sky ($\mu_V\approx 21.9$ mag arcsec$^{-2}$ in the $V$-band at the darkest telescope sites on Earth) is composed of several components: airglow, zodiacal light, light from the Milky Way, and the EBL. The latter, which consists of the integrated contributions from astronomical light sources over a wide range of redshifts, contributes only a small fraction to the overall night sky flux ($\mu_{\mathrm{EBL},V}\approx 25.4$ arcsec$^{-2}$, as estimated in section 3, or about 4\% of the total), but is the one component that is likely to originate largely from behind target objects in the low-redshift Universe.
\begin{figure*}
\includegraphics[scale=0.27]{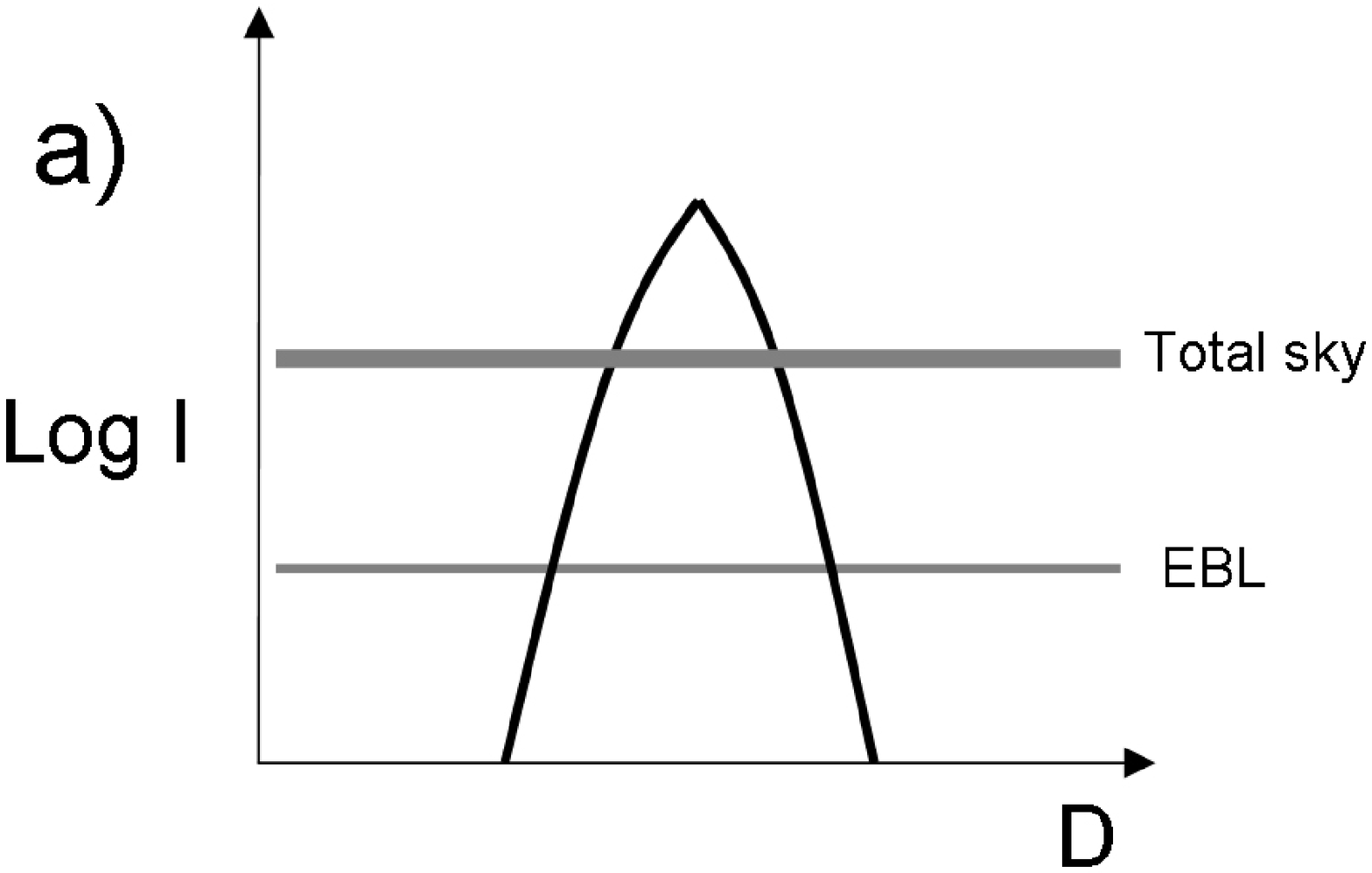}
\includegraphics[scale=0.27]{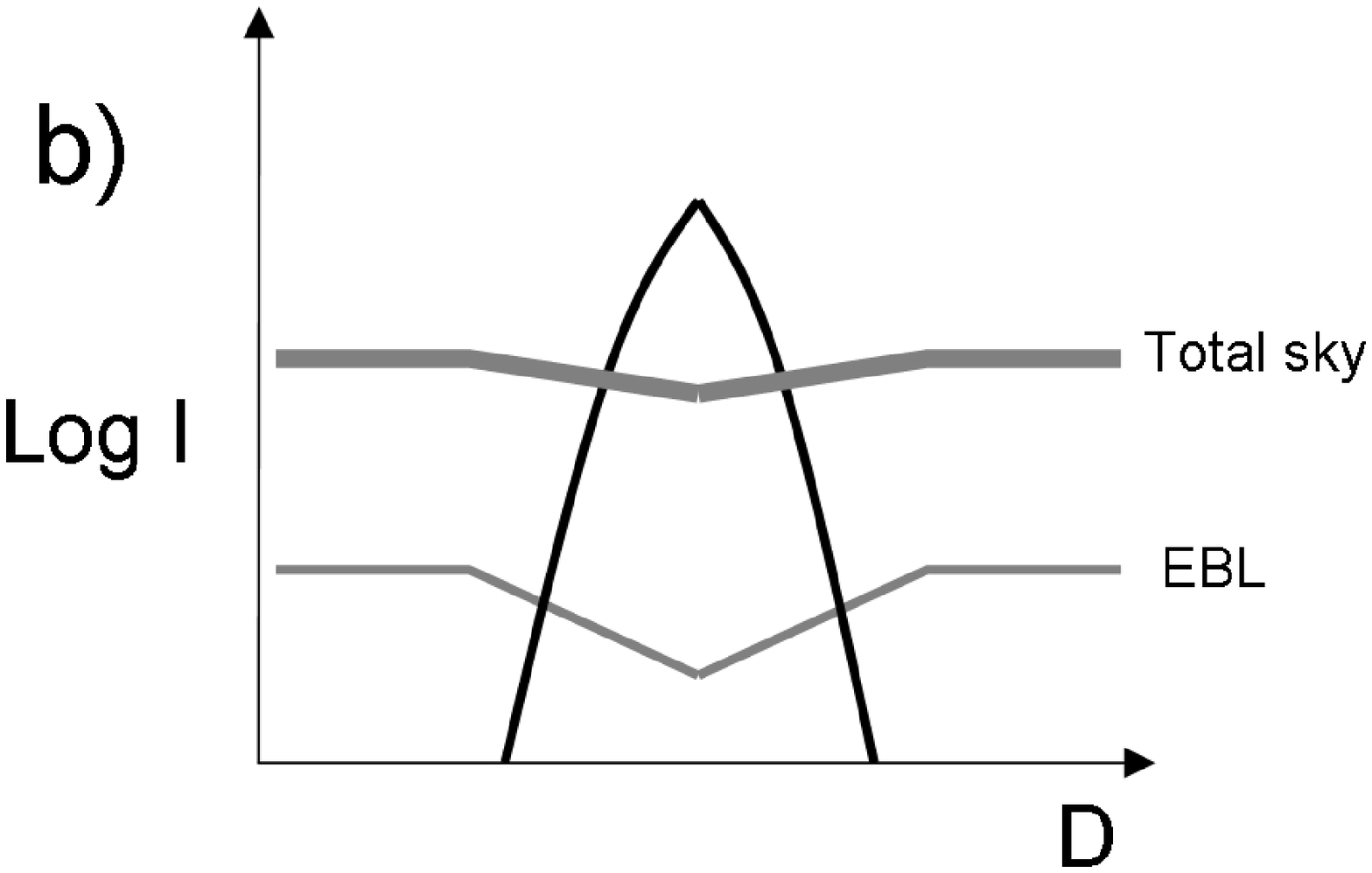}
\includegraphics[scale=0.27]{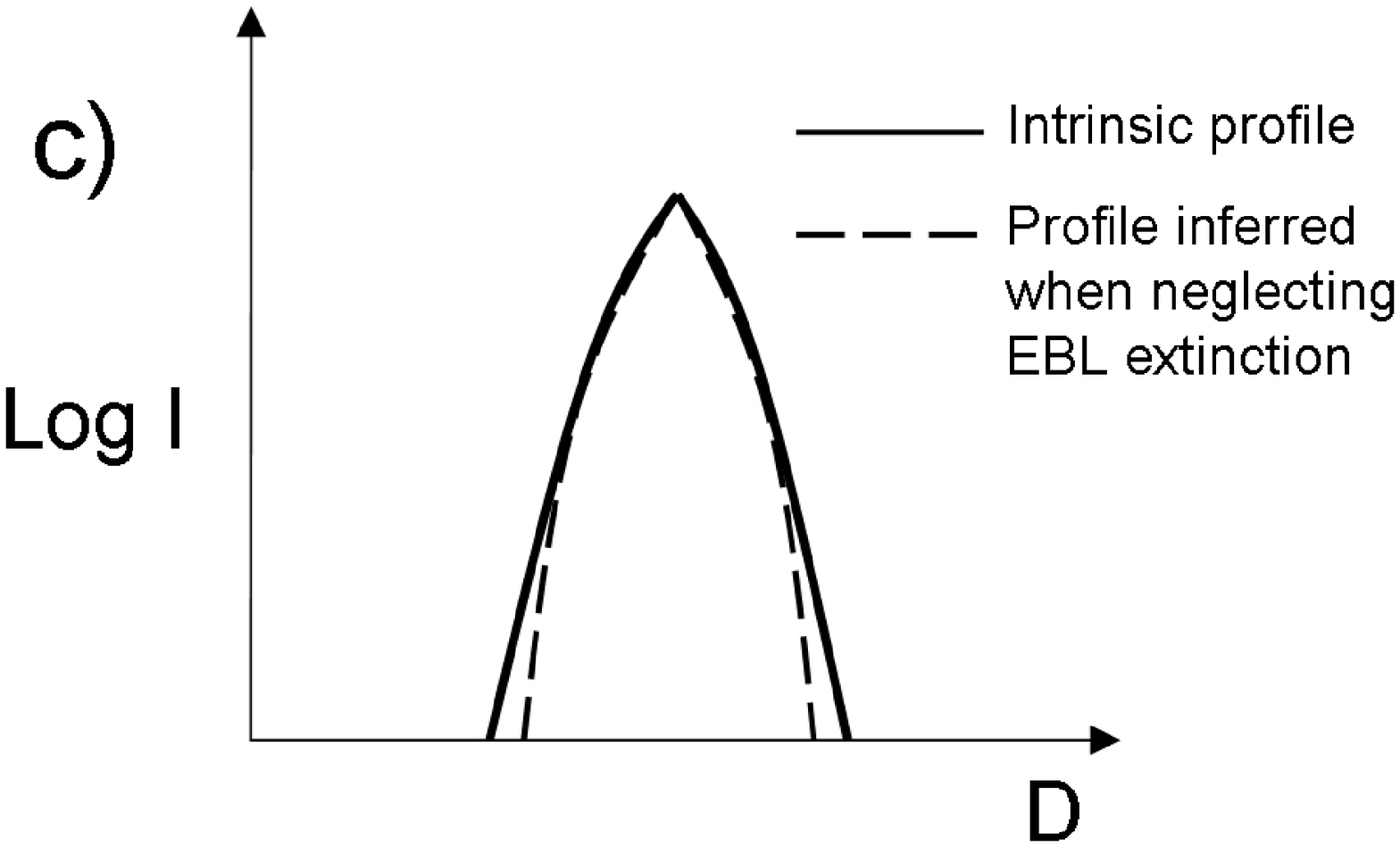}
\caption{Schematic illustration of the spurious effects introduced in surface brightness profiles by failure to take EBL extinction into account.
{\bf a)} Typical sky subtraction procedures assume that the EBL (thin gray line), and hence the night sky (thick gray line), is featureless across the surface brightness profile of an extended object (black line). {\bf b)} In reality, the surface brightness of the EBL may, due to extinction by dust present within the object, show a central depression across the face of the object, resulting in a central depression in the total surface brightness of the night sky. In relative terms, the depression in the total night sky is significantly weaker, due to dilution by other night sky components (airglow, zodiacal light, light from the Milky Way), which do not show any such depression. {\bf c)} As a result of assuming the situation depicted in {\bf a)}, rather than {\bf b)}, current sky subtraction procedures are unlikely to recover the intrinsic surface brightness profile of the object (solid line), but instead a profile (dashed line) for which the surface brightness drops too fast at large radii. This will not have any significant effect on the total fluxes and colours of bright objects, but may significantly affect studies of their low surface brightness regions.}
\label{fig2}
\end{figure*}

This turns out to be relevant when attempting to push surface photometry techniques to very faint surface brightness levels. To reach structures far below the surface brightness of the night sky, the subtraction of the sky flux needs to be as accurate as possible. As an example, surface photometry measurments at $\mu_V\approx 28.0$ mag arcsec$^{-2}$ with an uncertainty of $\pm 1.0$ mag require that the sky flux can be measured with an accuracy better than 99.9\%, which is a daunting task at the current time. 

In Fig.~\ref{fig1}, we illustrate how the sky flux is usually estimated from optical images. Consider the case of a galaxy, for which one would like to derive the surface brightness profile out to a radius marked by the dashed ellipse. One typically measures the sky flux in regions (here marked by boxes) well away from the target galaxy (at least outside the faintest isophote one aspires to reach) and other resolved sources in the frame. A surface or constant is then fitted to these measurements and subtracted from the frame. An implicit assumption in this procedure is that the sky level determined outside the target isophote (dashed ellipse) can somehow be interpolated into the regions where one attempts to measure the surface brightness of the galaxy studied. The problem is, that in the presence of dust within the target galaxy, the EBL component of the sky flux can be somewhat lower inside the galaxy than outside. Hence, the sky flux measured far away from the galaxy will systematically overestimate the sky flux at positions within the galaxy. 

The effect of this oversubtraction of sky flux on the resulting surface brightness profile of the galaxy is schematically illustrated in Fig.~\ref{fig2}. Fig.~\ref{fig2}a depicts the situation implicitly assumed when reducing astronomical images. The EBL (thin solid gray line) and consequently the total sky background (thick solid gray line) are incorrectly taken to be featureless across the surface brightness profile of the target object (solid black line). In Fig.~\ref{fig2}b, we depict what is likely to be a more realistic situation, in which the EBL displays a depression due to dust extinction in the direction towards the target object. As a result, the total sky background also shows a decrease in the direction towards the target. Fig.~\ref{fig2}c shows the resulting surface brightness profiles after sky subtraction. The incorrect assumption of a flat sky background across the object gives rise to an oversubtraction of sky, and hence a surface brightness profile which drops too fast at large distances from the centre (dashed black line) compared to the intrinsic profile (solid black line). Since the EBL constitutes such a small part of the overall sky flux, this problem will have negligible effects on the total and colours fluxes of bright galaxies, but may significantly affect surface photometry measurements of low surface brightness regions. One may argue that such regions are unlikely to contain much dust, but we will demonstrate that even tiny amounts of extinction can have significant effects at the faintest isophotes currently probed.

The sky subtraction situation just described matches that typically encountered when using images built from a stack of short exposures taken with small telescope offsets inbetween (a technique sometimes referred to as dithering). In the near-IR, one often removes the sky from dithered images by subtracting subsequent exposures from each other (provided that the offset has been greater than the diameter of the target isophote). In the optical -- and especially in the near-IR -- one sometimes also choses to move the telescope to one or several adjacent sky fields inbetween exposures of the target object (sky chopping). Either the entire sky frame, or a surface fitted to it, may then be subtracted from the object frame. However, these alternative sky subtraction techniques are all based on the same basic assumption as the method first outlined, namely that the sky flux outside the target object is representative for the sky flux inside. Therefore, all of these methods will result in a similar oversubtraction of the EBL.

The effects of oversubtracting the EBL can be estimated using a set of simple equations. Standard reduction techniques (which neglect the effects of dust on the EBL) implicitly assume that the observed surface brightness $I$ of the target (in units of $L_\odot \ \mathrm{pc}^{-2}$) at a certain wavelength $\lambda$ and position $r$ is given by:
\begin{eqnarray}
I_\mathrm{obj,\ approx}(\lambda,r) = I_\mathrm{obj}(\lambda,r) + I_\mathrm{airglow}(\lambda,r) + \nonumber\\ I_\mathrm{zodiacal}(\lambda,r) + I_\mathrm{MW}(\lambda,r) + I_\mathrm{EBL}(\lambda,r),
\label{eq1}
\end{eqnarray}
where $I_\mathrm{obj}$ is the intrinsic surface brightness of the target object and $I_\mathrm{airglow}$, $I_\mathrm{zodiacal}$, $I_\mathrm{MW}$ and $I_\mathrm{EBL}$ are the contributions from airglow, zodiacal light, Milky Way sources and EBL to the surface brightness of the night sky. However, since the EBL has passed through the dust layers of the target object, the observed surface brightness is more accurately given by: 
\begin{eqnarray}
I_\mathrm{obs}(\lambda,r) = I_\mathrm{obj}(\lambda,r) + I_\mathrm{airglow}(\lambda,r)+  I_\mathrm{zodiacal}(\lambda,r) +\nonumber\\ I_\mathrm{MW}(\lambda,r) + (1-f_\mathrm{abs}(\lambda,r))I_\mathrm{EBL}(\lambda,r),
\label{eq2}
\end{eqnarray}
where $f_\mathrm{abs}(\lambda,r)$ corresponds to the fraction of the EBL flux that is absorbed by dust at wavelength $\lambda$ and position $r$. Under the incorrect assumption that equation (\ref{eq1}) holds, standard sky subtraction methods would result in a surface brightness $I_\mathrm{obj,\ approx}(\lambda,r)$ for the target object:
\begin{eqnarray}
I_\mathrm{obj,\ approx} = I_\mathrm{obs}(\lambda,r) - I_\mathrm{airglow}(\lambda,r) -\nonumber\\ I_\mathrm{zodiacal}(\lambda,r) - I_\mathrm{MW}(\lambda,r) - I_\mathrm{EBL}(\lambda,r),
\label{eq3}
\end{eqnarray}
whereas the true, intrinsic surface brightness of the object is given by:
\begin{eqnarray}
I_\mathrm{obj}(\lambda,r) = I_\mathrm{obs}(\lambda,r) - I_\mathrm{airglow}(\lambda,r) -\nonumber\\ I_\mathrm{zodiacal}(\lambda,r) - I_\mathrm{MW}(\lambda,r) - (1-f_\mathrm{abs}(\lambda,r))I_\mathrm{EBL}(\lambda,r).
\label{eq4}
\end{eqnarray}

Hence, because of the incorrect assumptions going into equation (\ref{eq1}), standard sky subtraction methods introduce a relative error in the surface brightness of the target galaxy corresponding to:
\begin{eqnarray}
\frac{I_\mathrm{obj}(\lambda,r)-I_\mathrm{obj,\ approx}(\lambda,r)}{I_\mathrm{obj}(\lambda,r)} = \nonumber\\  \frac{f_\mathrm{abs}(\lambda,r)I_\mathrm{EBL}(\lambda,r)}{I_\mathrm{obj}(\lambda,r)} .
\label{photeq}
\end{eqnarray}

\begin{figure*}
\includegraphics[scale=0.44]{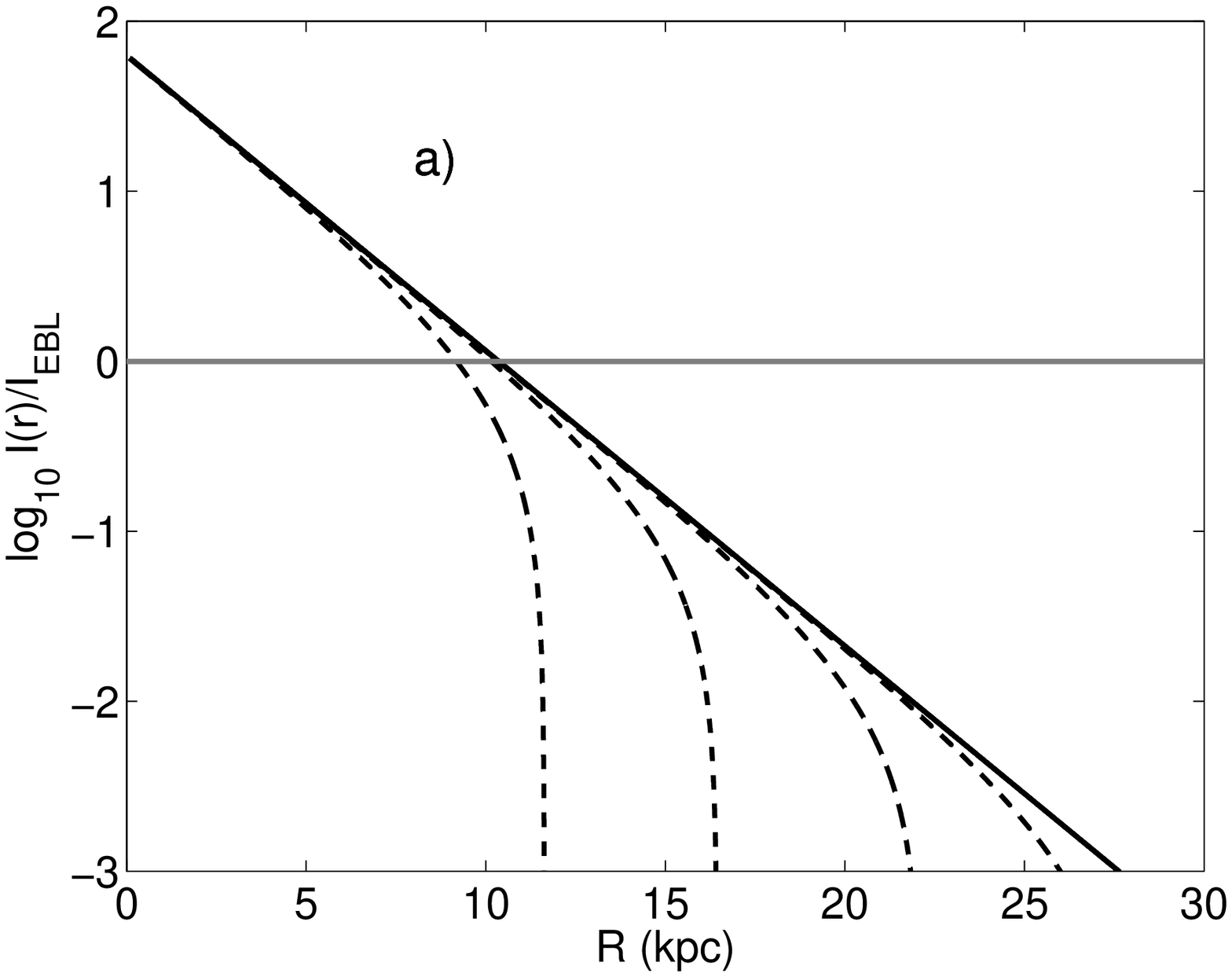}
\includegraphics[scale=0.44]{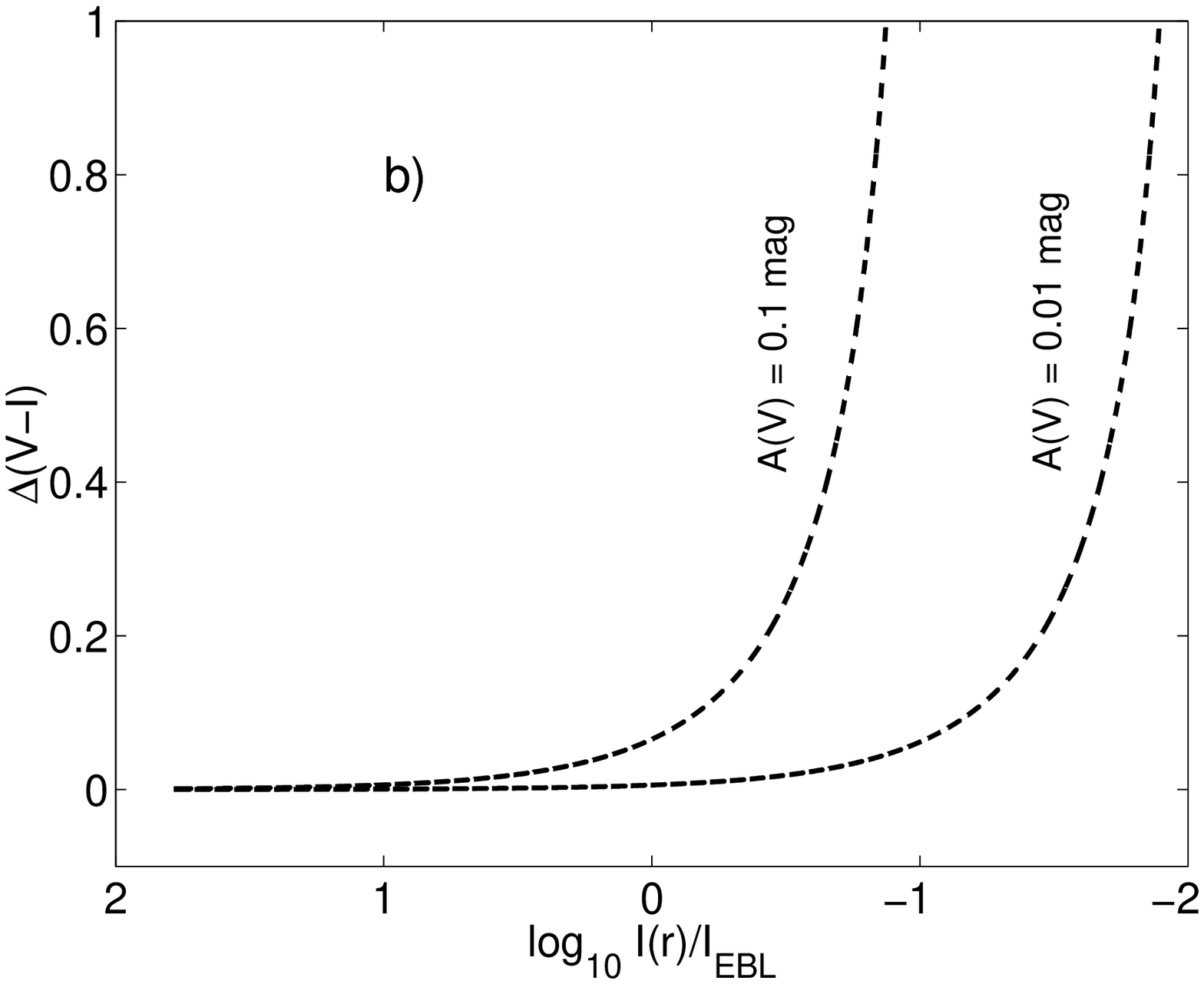}
\caption{The effects of spatially uniform EBL extinction (dust screen) on the $V$-band surface brightness profiles and colours of an exponential disc (central surface brightness $\mu_0=21$ mag arcsec$^{-2}$, scale length $h=2.5$ kpc, intrinsic colour $V-I=1.2$). {\bf a)} The intrinsic $V$-band surface brightness profile (black solid line) compared to the profiles resulting from failure to account for EBL extinction (dashed lines). The gray horizontal line represents the EBL level. The different dashed curves correspond to (from left to right at $\log_{10} I/I_{EBL}=-2.5)$ extinction values of $A(V)=1.0$, 0.1, 0.01 and 0.001 mag ($f_\mathrm{abs}(V)\approx 0.6$, 0.09, 0.009, 0.0009). The EBL extinction has pronounced effects on the profiles at surface brightness levels comparable to, or lower than that of the EBL itself. Even extinction values as low as $A(V)=0.01$ mag may give rise to detectable effects when pushing surface brightness measurements to current limits (e.g. $\log_{10} I/I_{EBL}\approx -2$, which corresponds to $\mu_V=30.5$ mag arcsec$^{-2}$ in the case of $\mu_{\mathrm{EBL},\ V}=25.5$ mag arcsec$^{-2}$ as derived in Section 3). {\bf b)} The spurious $\Delta(V-I)$ colour gradient introduced by EBL extinction in the case of $A(V)=0.1$ mag and 0.01 mag as a function of intrinsic $V$-band surface brightness. In the case of $A(V)=0.1$, significant colour offsets are introduced at $\log_{10} I/I_{EBL}< 0)$ (corresponding to $\mu_V>25.5$ mag arcsec$^{-2}$). For $A(V)=0.01$ mag, the same effect is seen at $\log_{10} I/I_{EBL}< -1$ ($\mu_V> 28.0$ mag arcsec$^{-2}$).}
\label{fig3}
\end{figure*}

For reasonable values on the dust absorption factor ($f_\mathrm{abs}\ll 1$), this error will be small when the surface brightness of the target object is greater that that of the EBL ($I_\mathrm{obj}(\lambda,r) \gg I_\mathrm{EBL}(\lambda,r)$). It may, however, become considerable when the surface brightness of the target becomes similar to, or lower, than the EBL ($I_\mathrm{obj}(\lambda,r) \leq I_\mathrm{EBL}(\lambda,r)$. Since the error introduced by the implicit assumptions going into equation (\ref{eq1}) results in a systematic oversubtraction of sky flux, this error is always positive. This means that the surface brightness of the target object will be systematically underestimated. In Fig.~\ref{fig3}a, we demonstrate this effect on exponential surface brightness profiles $I(r)=I_0\exp(-r/h)$ (often used to describe the surface brightness profile of disc galaxies) in the case of several different spatially constant dust absorption factor $f_\mathrm{abs}(\lambda)$ (see Section 4 for a treatment of more realistic disc opacity profiles). The deviation of the observed surface brightness profile from the intrinsic one becomes significant around $I_\mathrm{obj}\approx I_\mathrm{EBL}$, and keeps getting increasingly serious at fainter surface brightness levels. Once $f_\mathrm{abs}(\lambda,r)I_\mathrm{EBL}(\lambda,r)>I_\mathrm{obj}$, $I_\mathrm{obj,\ approx}$ will become negative, at which point the observer will supposedly assume that the signal has been lost in the sky noise.
 
As seen in Fig.~\ref{fig3}a, even an EBL extinction as low as $A(V)=0.01$ mag (corresponding to $f_\mathrm{abs}(V)\approx 0.009$) may give rise to detectable effects when pushing surface brightness measurements to their current limits ($\log_{10} I(r)/I_\mathrm{EBL}\leq -2$, corresponding to $\mu_V\geq 30.5$ mag arcsec$^{-2}$).

The discrepancy between the true and inferred surface brightness has a complicated wavelength dependence, which may introduce spurious colour gradients across the face of extended objects. In the case of a colour index $m_{\lambda,1}-m_{\lambda,2}$ (e.g. $V-I$, measured in magnitudes), the failure to account for EBL extinction will introduce a shift of  
\begin{eqnarray}
& \Delta(m_{\lambda_1}-m_{\lambda_2})= -2.5\log_{10}  \nonumber \\ 
& \left[\left( \frac{ I_\mathrm{obj}(\lambda_2,r)- f_\mathrm{abs}(\lambda_2,r)I_\mathrm{EBL}(\lambda_2,r) } {I_\mathrm{obj}(\lambda_1,r)-f_\mathrm{abs}(\lambda_1,r)I_\mathrm{EBL}(\lambda_1,r) }\right)  \frac{I_\mathrm{obj}(\lambda_1,r)}{I_\mathrm{obj}(\lambda_2,r)}\right]. 
\label{coloureq}
\end{eqnarray}

We stress that this colour shift is different from normal dust reddening, because of its dependence on the ratio $I_\mathrm{obj}/I_\mathrm{EBL}$. In principle, the EBL effect can shift the colour either redward or blueward depending on the extinction law, the spectrum of the target object and the spectrum of the EBL itself (see Sections 5 and 6 for examples). 

In Fig.~\ref{fig3}b, we use the values estimated in section 3 for the surface brightness of the diffuse EBL in filters $V$ and $I$ to demonstrate the spurious colour gradients introduced in $V-I$ across an exponential disc when EBL extinction is neglected. At surface brightness levels comparable to or lower than that of the EBL, the observed $V-I$ colour is driven away from the intrinsic value, eventually reaching values not expected for normal stellar populations. 

\section{Surface brightness of the optical/near-IR EBL}  
As illustrated in Fig.~\ref{fig1}, it is customary to mask resolved objects (foreground stars and background sources) before estimating the sky level to be subtracted from optical/near-IR images. As resolved background objects make up a fraction of the EBL, the part of the EBL that is relevant for EBL extinction effects is the {\it diffuse} component, i.e. the fraction of the EBL that remains after masking. We stress, however, that this component need not be diffuse in any absolute sense -- the only requirement is that the sources responsible for the diffuse EBL remain unresolved (and unmasked) in the image analyzed. Deep, high-resolution images will therefore contain a diffuse EBL smaller than that in shallow images taken under poor seeing conditions.

The surface brightness of the total EBL at optical/near-IR wavelengths remains very uncertain, mainly due to the problem of correcting for zodiacal light \citep*[e.g.][]{Dwek et al.,Matsumoto et al.,Mattila,Thompson et al.,Bernstein}. In what follows, we adopt the direct EBL measurements presented by \citet{Bernstein} in the {\it Hubble Space Telescope (HST)} WFPC2 filters F300W, F555W and F814W filter (similar to Johnson-Cousins $U$, $V$ and $I$). In the near-IR, we adopt the $J$-band measurement of \citet{Wright} and the $K$ and $L$ band measurements of \citet*{Gorjian et al.}. By linearly interpolating the EBL spectrum between the measured data points, we estimate the surface brightness of the total EBL in Johnson/Cousins $UBVRIJHK$ filters and SDSS $ugriz$ filters. The result is presented in the second column of Table 1. While EBL measurements are often presented in units of ergs $^{-1}$ cm$^{-2}$ sr$^{-1}$ \AA$^{-1}$ or similar, we here present all estimates in units of mag arcsec$^{-2}$, which are more commonly used in the field of surface photometry. Vega magnitudes are used for $UBVRIJHK$ and SDSS AB-magnitudes for $ugriz$. The errorbars presented simply reflect the errors quoted for the original EBL measurements, and do not take into account the errors introduced by the interpolation scheme. As discussed by \citet{Matsumoto et al.}, the direct EBL measurements hint at a peak in the EBL spectrum around the $J$ band (but see \citealt{Mattila} and \citealt{Thompson et al.} for different views), and the unknown shape of this putative feature makes the $z$-band (interpolation from $I$ to $J$) estimates for the EBL rather uncertain.

While some of the EBL implied by these measurements can be attributed to resolved galaxies, most of it remains unresolved even in the deepest {\it HST} images \citep{Pozzetti & Madau,Totani et al.}. Indirect constraints based on the opacity of the Universe at gamma-ray wavelengths \citep*[e.g.][]{Franceschini et al.,Mazin & Raue,Albert et al.,Razzaque et al.} tend to favour an EBL level closer to that attributable to the resolved sources, but such methods may have problems of their own \citep[e.g.][]{Krennrich et al.,De Angelis et al.}. To derive the surface brightness of the resolved EBL, we adopt the estimates computed by \citet{Madau & Pozzetti} by integrating {\it HST} galaxy counts in a set of 7 broadband filters (somewhat similar to Johnson-Cousins $UBRIJHK$) from the Hubble Deep Fields. These galaxy counts are formally valid to faint AB magnitude limits varying from 30.5 mag at 6700 \AA \ to 25.5 mag at 22000 \AA, but \citet{Madau & Pozzetti} argue that the EBL contribution from resolved sources has converged in all the filters considered. We caution, however, that alternative methods for estimating the resolved component of the EBL give somewhat different results \citep[compare][]{Madau & Pozzetti,Totani et al.,Bernstein et al. a}. By interpolating the \citet{Madau & Pozzetti} estimates, we arrive at the $UBVRIJHK$ and $ugriz$ estimates for the surface brightness of the resolved EBL presented in the third column of Table 1.

To assess the surface brightness of the diffuse (unresolved) EBL, we subtract the resolved EBL component from the total:
\begin{equation}
I_\mathrm{EBL,\ diffuse}(\lambda)=I_\mathrm{EBL,\ total}(\lambda)-I_\mathrm{EBL,\ resolved}(\lambda).
\label{DiffuseEBLeq}
\end{equation} 
Here, we for simplicity assume that the all of the EBL that can currently be attributed to resolved galaxies has been resolved and masked in the images on which surface photometry is attempted. While this is likely to hold only for deep images taken from space or under excellent seeing conditions from the ground, the corrections due to magnitude limits different from those adopted above can be assessed from \citet{Pozzetti & Madau}. Under the assumptions that the total EBL measurements cited above do not contain any major systematic errors, corrections of this type turn out to be modest in all filters considered. The resulting surface brightness of the diffuse EBL are presented in the fourth column of Table 1. We make no attempt to assess the error bars on these entries, since systematic effects are likely to dominate the uncertainties. 
\begin{table*}
\caption{The estimated surface brightness of the total, resolved and diffuse EBL in Johnson-Cousins $UBVRIJHK$ and SDSS $ugriz$ filters. The surface brightness of the diffuse component is obtained by subtracting the EBL attributable to currently resolved galaxies from the total EBL. $UBVRIJHK$ magnitudes are in the Vega photometric system and $ugriz$ magnitudes in the SDSS AB system. See main text for additional details.}
  \begin{tabular}{@{}lllll@{}}
  \hline
Filter  & Total EBL & Resolved EBL & Diffuse EBL\\
        & (mag arcsec$^{-2}$) & (mag arcsec$^{-2}$) & (mag arcsec$^{-2}$)\\
\hline
$U$			& $26.0\ ^{+0.9}_{-0.5}$ & $28.4\ ^{+0.2}_{-0.2}$ & 26.1\\
$B$			& $26.1\ ^{+0.8}_{-0.4}$ & $28.6\ ^{+0.2}_{-0.1}$ & 26.3\\
$V$			& $25.4\ ^{+0.7}_{-0.4}$ & $27.9\ ^{+0.2}_{-0.1}$ & 25.5\\
$R$			& $25.0\ ^{+0.8}_{-0.4}$ & $27.4\ ^{+0.2}_{-0.1}$ & 25.1\\
$I$			& $24.6\ ^{+0.9}_{-0.5}$ & $26.7\ ^{+0.2}_{-0.1}$ & 24.7\\
$J$			& $24.3\ ^{+0.5}_{-0.4}$ & $25.5\ ^{+0.2}_{-0.2}$ & 24.7\\
$H$			& $23.9\ ^{+0.3}_{-0.3}$ & $25.1\ ^{+0.3}_{-0.2}$ & 24.4\\
$K$			& $23.1\ ^{+0.3}_{-0.3}$ & $24.2\ ^{+0.2}_{-0.2}$ & 23.5\\
$u$			& $26.7\ ^{+0.9}_{-0.5}$ & $29.1\ ^{+0.2}_{-0.2}$ & 26.8\\
$g$			& $25.8\ ^{+0.7}_{-0.4}$ & $28.3\ ^{+0.2}_{-0.1}$ & 25.9\\
$r$			& $25.2\ ^{+0.7}_{-0.4}$ & $27.7\ ^{+0.2}_{-0.1}$ & 25.4\\
$i$			& $25.0\ ^{+0.8}_{-0.5}$ & $27.3\ ^{+0.2}_{-0.1}$ & 25.2\\
$z$			& $24.9\ ^{+1.0}_{-0.5}$ & $26.9\ ^{+0.2}_{-0.1}$ & 25.1\\
\hline
\end{tabular}
\end{table*}

Inspection of the entries in Table 1 reveals that the surface brightness of the total EBL lies in the range $\mu\approx 25$--26 mag arcsec$^{-2}$ in the optical ($UBVRI$) and $\mu\approx 23$--25 mag in the near-IR ($JHK$). This is considerably brighter than the isophotal levels typically probed through optical surface photometry, and around the limit of what ground-based near-IR surface photometry can currently reach. Since the resolved galaxy population seems to make up no more than $\approx 1/10$ (optical) to $\approx 1/3$ (near-IR) of the total EBL, the surface brightness of the diffuse EBL is only 0.1--0.5 mag fainter than the total.

\begin{figure*}
\includegraphics[scale=0.45]{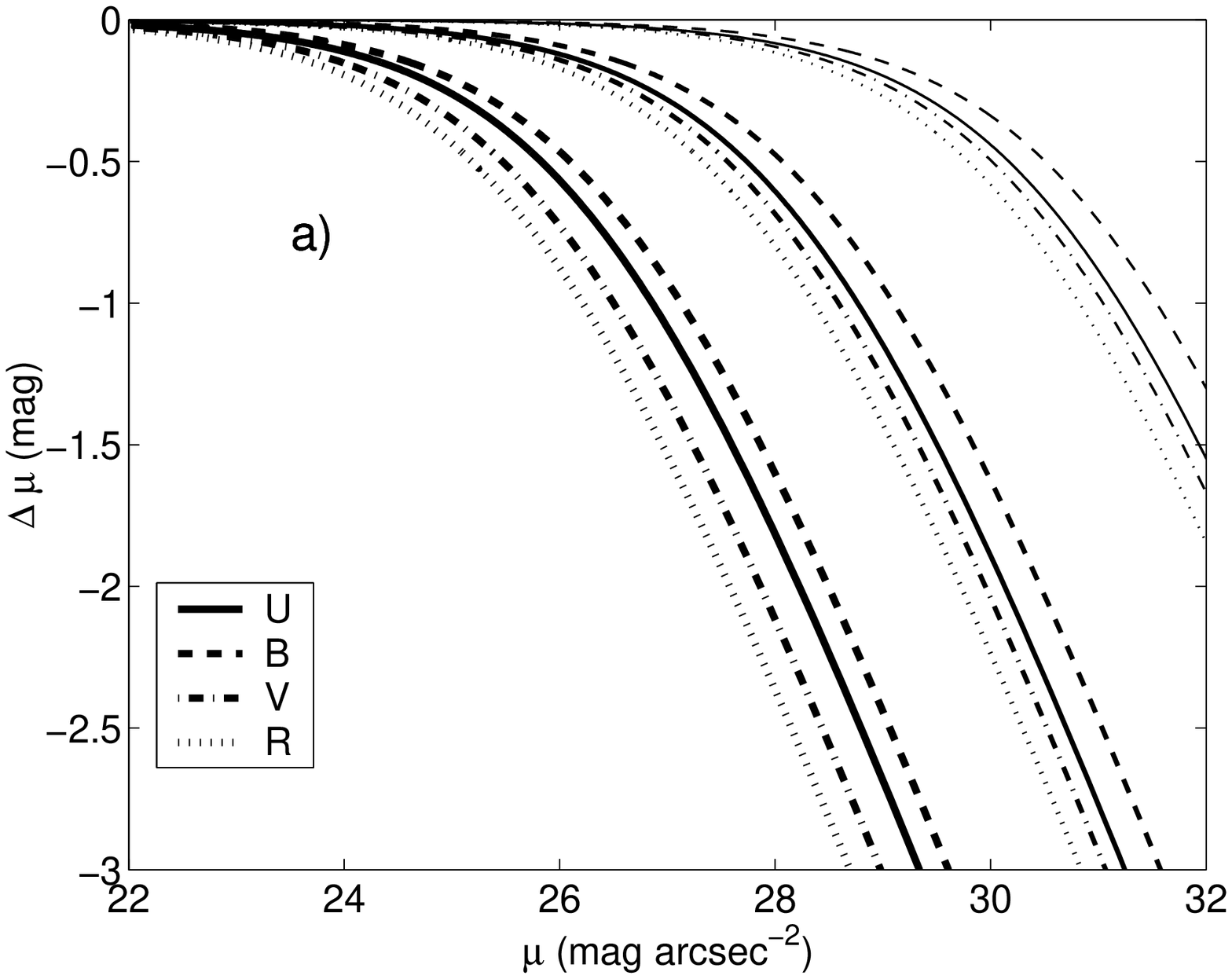}
\includegraphics[scale=0.45]{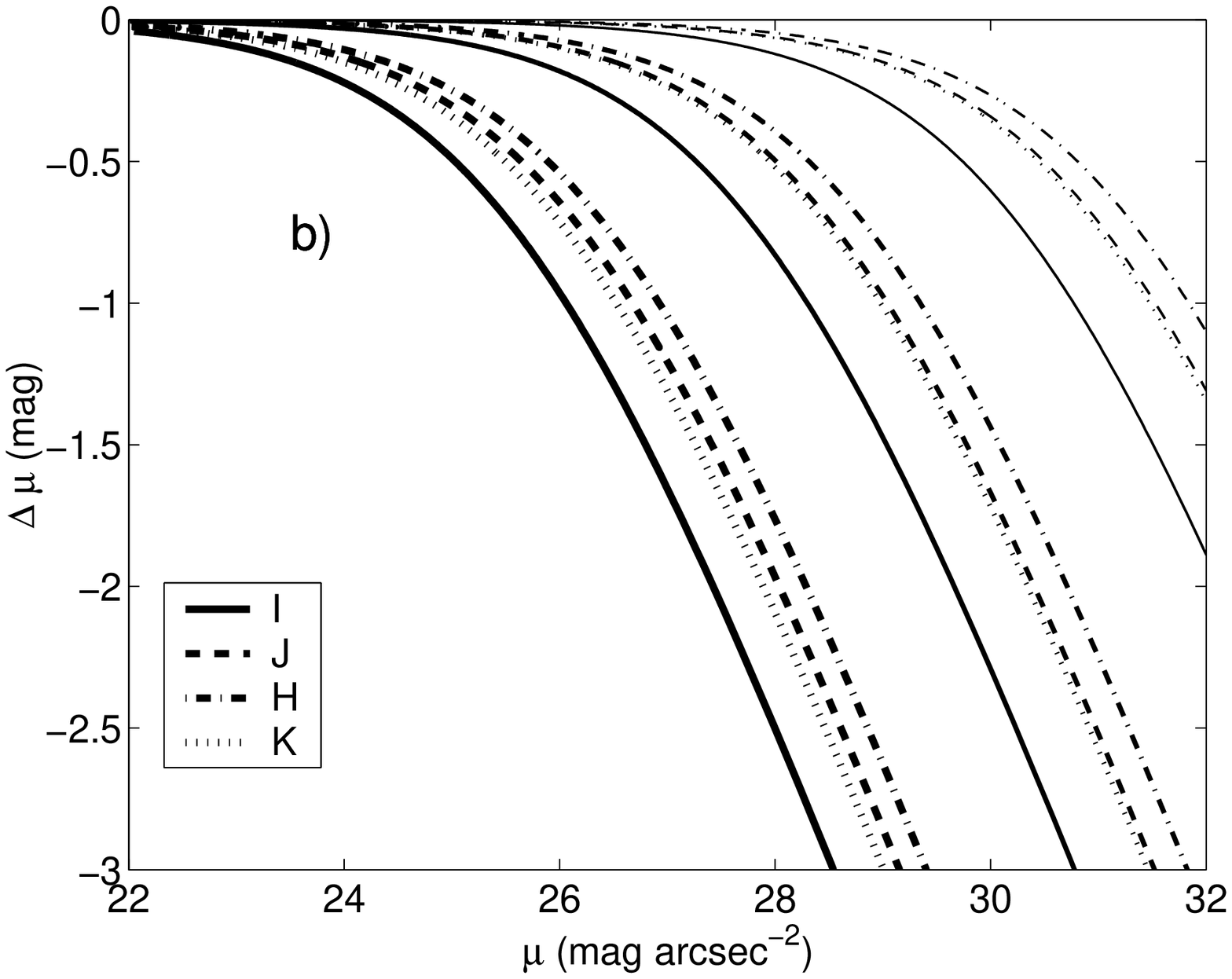}
\caption{The predicted offset $\Delta\mu$ between the intrinsic and observed surface brightness, as a function of observed surface brightness $\mu$ in different filters for $A(V)=1.0$ mag (thick lines), $A(V)=0.1$ mag (medium lines) and $A(V)=0.01$ mag (thin lines). A Milky Way extinction curve has been assumed. {\bf a)} Filters $U$ (solid), $B$ (dashed), $V$ (dash-doted) and $R$ (dotted). {\bf b)} Filters $I$ (solid), $J$ (dashed), $H$ (dash-dotted) and $K$ (dotted).}
\label{fig4}
\end{figure*}

Given that EBL extinction is likely to become relevant at surface brightness levels similar to or fainter than that of the diffuse EBL (see Fig.~\ref{fig3}), one may suspect this mechanism to impose strong limits on how deep surface photometry measurements can be pushed without the introduction of systematic errors. In Fig.~\ref{fig4}, we show the predicted offset $\Delta\mu$ between the intrinsic and observed surface brightness, as a function of {\it observed} surface brightness $\mu_\mathrm{obj,\ approx}$ (i.e. the values inferred when EBL extinction is neglected), in filters $UBVRIJHK$ for spatially constant $V$-band extinction values of $A(V)=1.0$ mag (thick lines), 0.1 mag (medium lines) and 0.01 mag (thin lines), assuming a Milky Way extinction curve. These offset predictions are independent of the surface brightness profile of the target object (as long as the extinction is considered spatially constant), and can be used to give a rough assessment of the likely impact of EBL extinction on any set of surface photometry data. The offsets turn out to be fairly independent of the filters used. This ``conspiracy'' stems from the fact that the brighter EBL in the red part of the spectrum (when measured in Vega magnitudes) is compensated for by the lower extinction given by Milky Way extinction at those wavelengths. 

Systematic errors of $\Delta\mu\approx 1$ mag are produced at $\mu\approx 27$ mag arcsec$^{-2}$ for $A(V)=1.0$ mag, at $\mu\approx 29$ mag arcsec$^{-2}$ for $A(V)=0.1$ mag and at $\mu\approx 31.5$ mag arcsec$^{-2}$ for $A(V)=0.01$ mag. Significant EBL effects are therefore expected in current optical surface photometry data (which can reach as deep as $\approx 32$ mag arcsec$^{-2}$), whereas only modest effects are expected in the near-IR bands ($JHK$), since current observations rarely probe beyond $\mu\approx 23$--24 mag arcsec$^{-2}$ at these wavelengths.

\section{Impact on studies of disc truncations}
Fig.~\ref{fig3} suggests that surface brightness, when studied through deep surface photometry, should fall faster than that for an exponential in the outer, faint regions of discs. Such truncated surface brightness profiles are in fact commonly observed among disc galaxies \citep[for a recent review, see][]{Pohlen et al.}, and one might ask whether some fraction of these features may be due to EBL extinction effects rather than intrinsic disc truncations. The possibility that artificial disc truncations may arise from oversubtraction of sky has previously been discussed by \citet{Barteldrees & Dettmar} \citet{de Grijs et al.} and \citet{Narayan & Jog}, but the effect discussed here is markedly different, since the amount of oversubtraction associated with the EBL is a function of both intrinsic disc surface brightness and dust content. Observationally, the truncation radius of both high and low surface brightness discs appears at an average $I$-band surface brightness of $\mu_I\approx 25.3$ mag arcsec$^{-2}$ \citep{Kregel & van der Kruit}. Inspection of Table 1 reveals that this is fainter than the diffuse EBL level inferred by direct measurements ($\mu_{I}\approx 24.7$ mag arcsec$^{-2}$), and hence in the surface brightness regime where EBL extinction effects are expected to turn up. 

To assess the magnitude of the EBL effect in the outskirts of discs, we adopt a model for radial disc opacity $\tau(r)$ of the type: 
\begin{equation}
\tau(r)=\tau(0)\exp\left(-\frac{r}{h_\tau}\right),
\label{eq6}
\end{equation}
where $\tau(0)$ is the central opacity and $h_\tau$ is an extinction scale length. This model has been demonstrated to give a reasonable fit to the observed opacity distribution, as derived both from occulting galaxy pairs and counts of background galaxies through foreground discs \citep*[e.g][]{Holwerda et al. a,Holwerda et al. b}. In what follows, we adopt $\tau_0=0.2$ and $h_\tau=2.9 R_\mathrm{eff}$ (where $R_\mathrm{eff}$ is the effective radius of the disc), as derived for $0.01<z<0.1$ galaxies in the SDSS by \citet{Holwerda et al. b}. This opacity profile was derived by stacking data in the SDSS $r$ and $i$ filters. To derive the opacity at other wavelengths, we assume the original opacity profile to be valid at $\lambda\approx 6900$ \AA{ } and simply rescale $\tau_0$ by assuming either a Milky Way extinction curve \citep{Pei} or gray extinction. The latter option is supported by absence of any obvious trend between the colours of background galaxies seen through foreground discs and the inferred disc opacity \citep{Holwerda et al. a} and can be attributed to a clumpy interstellar medium \citep{Keel & White}. 

Using the relation $R_\mathrm{eff}\approx 1.68h$ for an infinite exponential disc, we get a ratio between the opacity scalelength and the scalelength of the luminous disc of $h_\tau/h\approx 5$. 

\begin{figure*}
\includegraphics[scale=0.45]{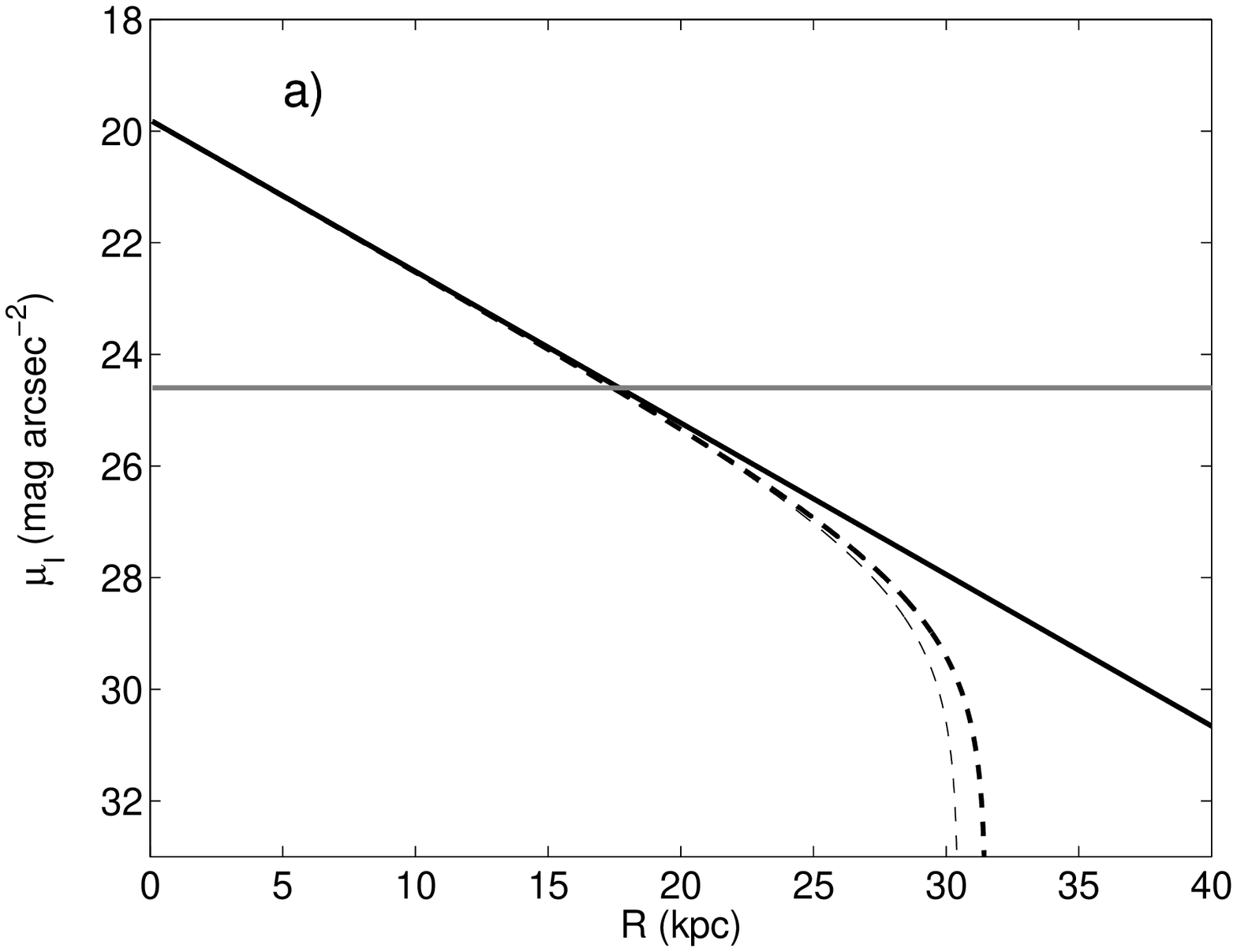}
\includegraphics[scale=0.45]{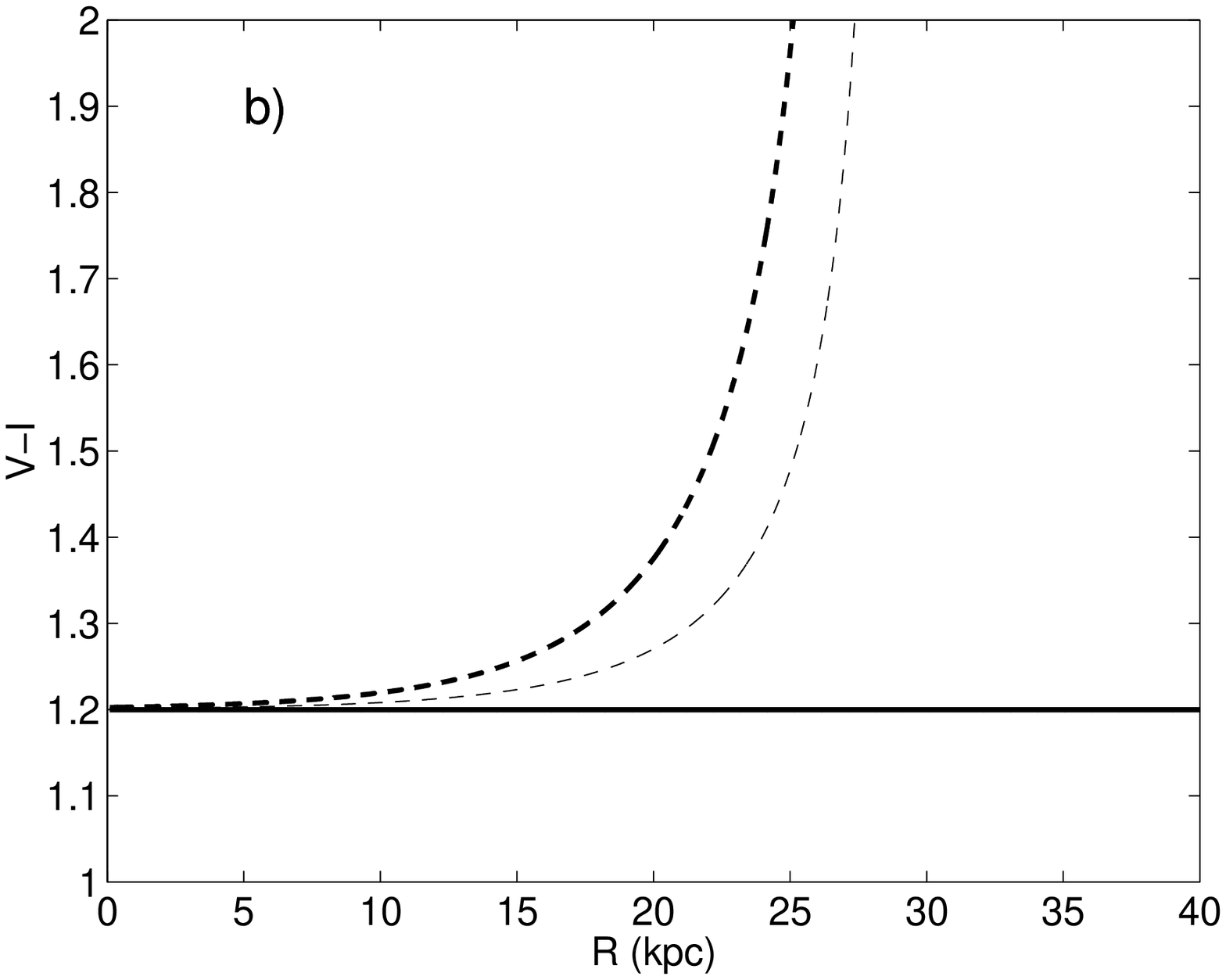}
\caption{The predicted effects of EBL extinction on an exponential disc with central surface brightness $\mu_{V,0}=21$ mag arcsec$^{-2}$, scale length $h=4$ kpc, an intrinsic colour of $V-I=1.2$ at all radii and opacity properties as specified in the main text. {\bf a)} The intrinsic $I$-band surface brightness profile (thick solid) compared to the observed profiles predicted for a Milky Way extinction curve (thick dashed) and gray extinction (thin dashed). The horizontal, gray solid lines indicate the EBL level in this filter. While the EBL extinction effect mimics a soft disc truncation, the change in profile slope occurs at much fainter isophotes ($\mu_I\approx 27$ mag arcsec$^{-2}$) than for typical disc galaxies ($\mu_I\approx 25.3$ mag arcsec$^{-2}$; \citealt{Kregel & van der Kruit}). {\bf b)} The intrinsic $V-I$ colour profile (thick solid) compared to the predicted colour profile in the case of a Milky Way extinction curve (thick dashed) and gray extinction (thin dashed). Due to a more rapid decrease in $V$-band flux at 25--30 kpc from the centre compared to the $I$-band, $V-I$ is driven to extremely high values at these radii.}
\label{fig5}
\end{figure*}

In Fig.~\ref{fig5}, we demonstrate the effects of EBL extinction on a typical disc galaxy with $\mu_{0,V}=21.0$ mag arcsec$^{-2}$, $h=4$ kpc \citep*{MacArthur et al.} in filters $V$ and $I$, assuming intrinsic colours of $V-I=1.2$, and no population gradient. As seen in Fig.~\ref{fig5}a, the EBL extinction gives rise to a smooth change in slope of the outer surface brightness profile (at around $r\geq 25$--30 kpc or $\approx 6$--8 scalelengths from the centre). At this distance from the centre, the extinction is predicted to be around 0.10--0.05 magnitudes (depending on the extinction law) in the $V$-band and around 0.05 magnitudes in the $I$-band. This break resembles the truncations commonly seen in the outer parts of disc galaxies, but here occurs around $\mu_I\approx 27$ mag arcsec$^{-2}$ compared to the observed value of $\mu_I\approx 25.3$ mag arcsec$^{-2}$ \citep{Kregel & van der Kruit}. The choice of extinction law has but a minor impact on the appearance of this feature, as evident from the small difference between the thick dashed line (Milky Way extinction) and thin dashed line (gray extinction) in Fig.~\ref{fig5}a. This implies that the EBL extinction effect cannot be the primary cause of the observed disc truncations. Since the rapid decrease in surface brightness caused by EBL extinction is predicted to set in at slightly different radii in the $V$ and $I$ bands, a strong, spurious colour gradient will develop in the outskirts of the disc (Fig.~\ref{fig5}b). At $r\approx 25$--30 kpc, $V-I$ is suddenly driven from the intrinsic value of $V-I=1.2$ to $V-I>1.6$, which is not expected for normal stellar populations, even at high metallicities. 

\begin{figure}
\includegraphics[width=84mm]{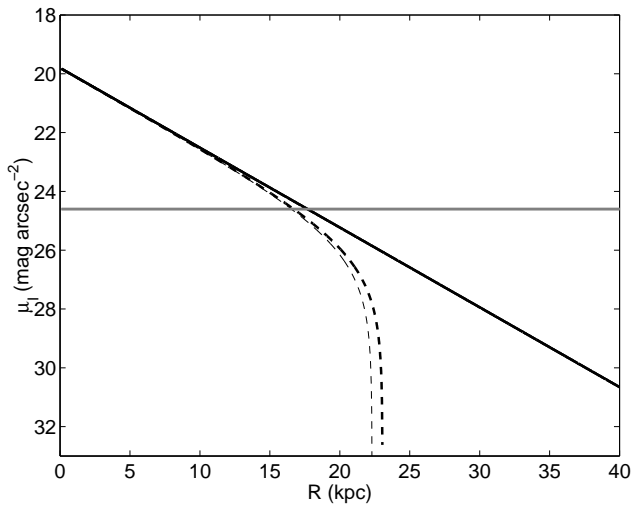}
\caption{Same as Fig.~\ref{fig5}a, but for a disc with $\tau=0.5$ and $h_\tau/h=20$ instead of  instead of $\tau=0.2$ and $h_\tau/h=5$. The boosted opacity of this disc shifts the drop in the surface brightness profile to brighter isophotes ($\mu_I\approx 25$ mag arcsec$^{-2}$ instead of $\mu_I\approx 27$ mag arcsec$^{-2}$ for the more transparent disc adopted in Fig.~\ref{fig5}a) and smaller radii.}
\label{fig6}
\end{figure}

While EBL extinction is unlikely to be the main cause of the observed disc truncations, the agreement between the observed and predicted disc breaks improves for dusty discs. This is demonstrated in Fig.~\ref{fig6}, where the ratio between the opacity scalelength and the scalelength of the luminous disc has been increased to $h_\tau/h\approx 20$ \citep[in agreement with the average value found for the smaller sample presented by][]{Holwerda et al. a} and the central opacity has been boosted to $\tau_0=0.5$. This places the spurious break in the $I$-band profile at around $\mu_I\approx 25$ mag arcsec$^{-2}$, in reasonable agreement with the break radius typically observed. 

This opacity model gives a central $V$-band EBL extinction of $A(V)\approx 0.7$ (0.55) mag and $A(V)\approx 0.55$ (0.4) mag at 25 kpc from the centre for a Milky Way (gray) extinction  curve. Assuming an intrinsic $B-V$ colour of $B-V=0.6$, this distance corresponds to $r\approx 1.5\times R_{25}$ for the model disc, where $R_{25}$ is the radius of the $\mu_B=25$ mag arcsec$^{-2}$ isophote. Inspection of the opacity profiles derived in \citet{Holwerda et al. a} and \citet{Holwerda et al. b} reveals that an opacity model of this type may well be consistent with some of the more opaque disc galaxies in their samples (although the error bars are typically very large for individual galaxies at these radii). This implies that while EBL extinction is unlikely to be the main reason for the observed disc truncations, some of the truncations seen in unusually dusty discs may possibly be attributed to this mechanism. While disc truncations often manifest themselves as a break between two exponential profiles with different scale lengths (in plots with $\mu$ versus $R$ such as Fig.~\ref{fig5}a and Fig.~\ref{fig6}, this would correspond to a straight line which suddenly changes slope at the break radius), EBL extinction effects predict a surface brightness profile which turns progressively steeper and eventually becomes vertical (see Fig.~\ref{fig3}a, Fig.~\ref{fig5}a and Fig.~\ref{fig6}). The distinction may, however, be difficult to make in the presence of large observational errorbars at the fainter isophotes, and the fact that surface photometry of discs rarely probe regions beyond $\mu_I=28$ mag arcsec$^{-2}$. At this radius, the vertical part of the profile (where the surface brightness tends to zero due to oversubstraction of sky) has not yet been reached (see Fig.~\ref{fig5}a and  Fig.~\ref{fig6}). Moreover, any deviation from the smooth opacity profile given by equation (\ref{eq6}) will also distort the appearance of this feature.

A strong colour gradient in the outskirts of discs, like the one leading to unphysically high values for $V-I$ in Fig.~\ref{fig5}b, could in principle serve as a tell-tale signature of EBL extinction. However, inspection of equation (\ref{coloureq}) reveals that while similar colour gradients are expected for many combinations of EBL and galaxy spectra, they are by no means generic. As an example, consider the case where the spectrum of the EBL and the target galaxy have similar shape ($I_\mathrm{obj}(\lambda_1,r)/I_\mathrm{obj}(\lambda_2,r) \approx I_\mathrm{EBL}(\lambda_1,r)/I_\mathrm{EBL}(\lambda_2,r)$). In the presence of gray extinction ($f_\mathrm{abs}(\lambda_1,r)\approx f_\mathrm{abs}(\lambda_2,r)$), one gets $\Delta(m_{\lambda_1}-m_{\lambda_2})\approx 0$, leading to no detectable colour gradient. Hence, while strange colour gradients in the outer regions of discs may serve as an indication that the EBL extinction effect is at work, the absence of such gradients does not necessarily prove that such effects are unimportant. 
 
\section{Impact on studies of the halos of disc galaxies}
\begin{figure}
\includegraphics[width=84mm]{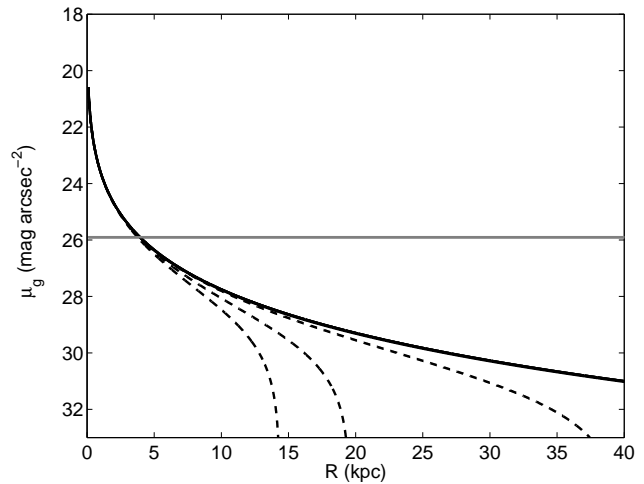}
\caption{The intrinsic $g$-band surface brightness profile (thick solid) compared to the observed profiles (thick dashed) predicted for a spatially constant extinction of $A(g)=0.1$, 0.05 or 0.01 mag (from left to right at $\mu_g=32$ mag arcsec$^{-2}$). The horizontal, gray solid line indicates the EBL level in this filter.}
\label{fig7}
\end{figure}
 
Attempts to probe the stellar halos of disc galaxies through surface photometry remains extremely challenging due to the very faint surface brightness levels at which such structures are expected to become dominant ($\mu_V\sim 29$ mag arcsec$^{-2}$; \citealt{Abadi et al.}; \citealt{Gauthier et al.}). Detections of such halos around edge-on disc galaxies have been claimed \citep[e.g.][]{Sackett et al.,Lequeux et al.,Rudy et al.,James & Casali,Zibetti et al. a,Zibetti & Ferguson,Caldwell & Bergvall}, but many of these remain controversial due to problems in subtracting the sky with sufficient accuarcy \citep{Zheng et al.} and correcting for instrumental scattering \citep{de Jong}. A common feature of the reported halos is a very red spectral energy distribution, manifesting itself as excess flux with respect to expectations for normal stellar populations in the wavelength range from $i$ ($\approx 7600$ \AA) to $K$ ($\approx 22000$ \AA). If correct, this could be indicative of a stellar halo with a very bottom-heavy initial mass function \citep{Lequeux et al.,Rudy et al.,Zackrisson et al. a}, with important implications for the missing-baryon problem \citep[e.g.][]{Zackrisson & Flynn}.

Figs.~\ref{fig3} and \ref{fig5} have already indicated that EBL extinction may produce artificial reddening of stellar populations at faint surface brightness levels. Hence, it becomes relevant to explore whether this could be the reason for the anomalous colours reported for the halos of disc galaxies. We here focus on the halo measurements made in the SDSS $gri$ filters (used by \citealt{Zibetti et al. a} and \citealt{Caldwell & Bergvall}). In Fig.~\ref{fig7}, we demonstrate the predicted effects of EBL extinction on the $g$-band data of a halo following the Sersic profile:
\begin{equation}
I=I_\mathrm{eff}\exp\left( -b_\mathrm{n}\left[(r/r_\mathrm{eff})^{1/n}-1 \right] \right),
\label{Sersiceq}
\end{equation}
with $b_\mathrm{n}\approx 2n-0.324$ \citep{Ciotti}, $n=7$, $r_\mathrm{eff}=10$ kpc. This intrinsic profile has been adopted for illustrative purposes only, with the values for $n$ and $r_\mathrm{eff}$ chosen to allow the surface photometry data of \citet{Zibetti et al. a} to be approximately reproduced, {\it after} EBL extinction effects have been introduced. However, our results concerning EBL extinction as the source of abnormal halo colours do not depend on the details of the intrinsic profile.

The thick solid line in Fig.~\ref{fig7} corresponds to the intrinsic surface brightness profile whereas the dashed lines indicate the profiles produced by a spatially constant extinction of $A(g)=0.1$, 0.05 or 0.01 mag (from left to right at $\mu_g=32$ mag arcsec$^{-2}$). With this amount of dust extinction, the observed surface brightness profile becomes severely affected a couple of magnitudes faintward of the diffuse EBL level (gray horizontal line), just like in the case of disc-like profiles (see Figs.~\ref{fig3} \& \ref{fig5}). 

\begin{figure*}
\includegraphics[scale=0.45]{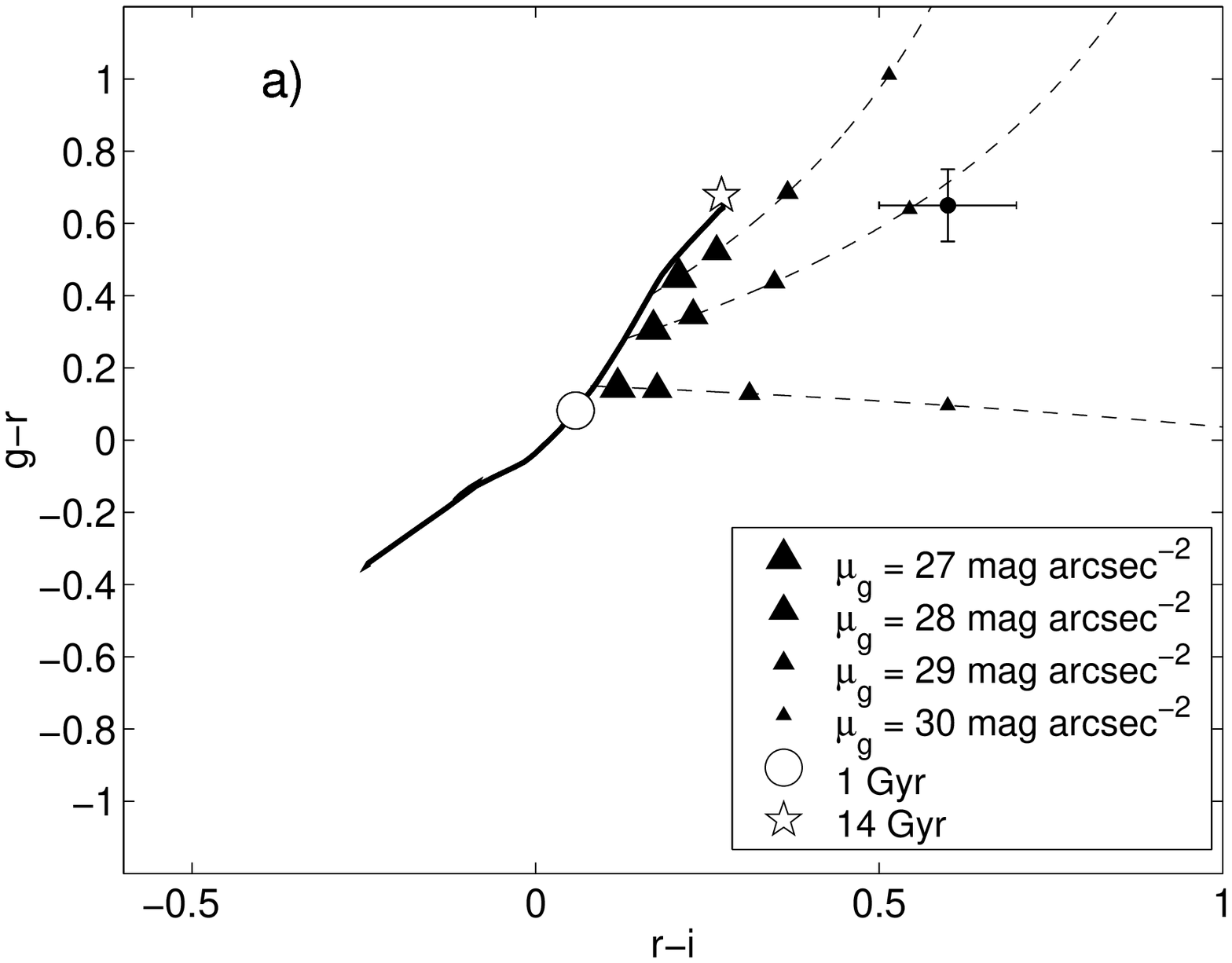}
\includegraphics[scale=0.45]{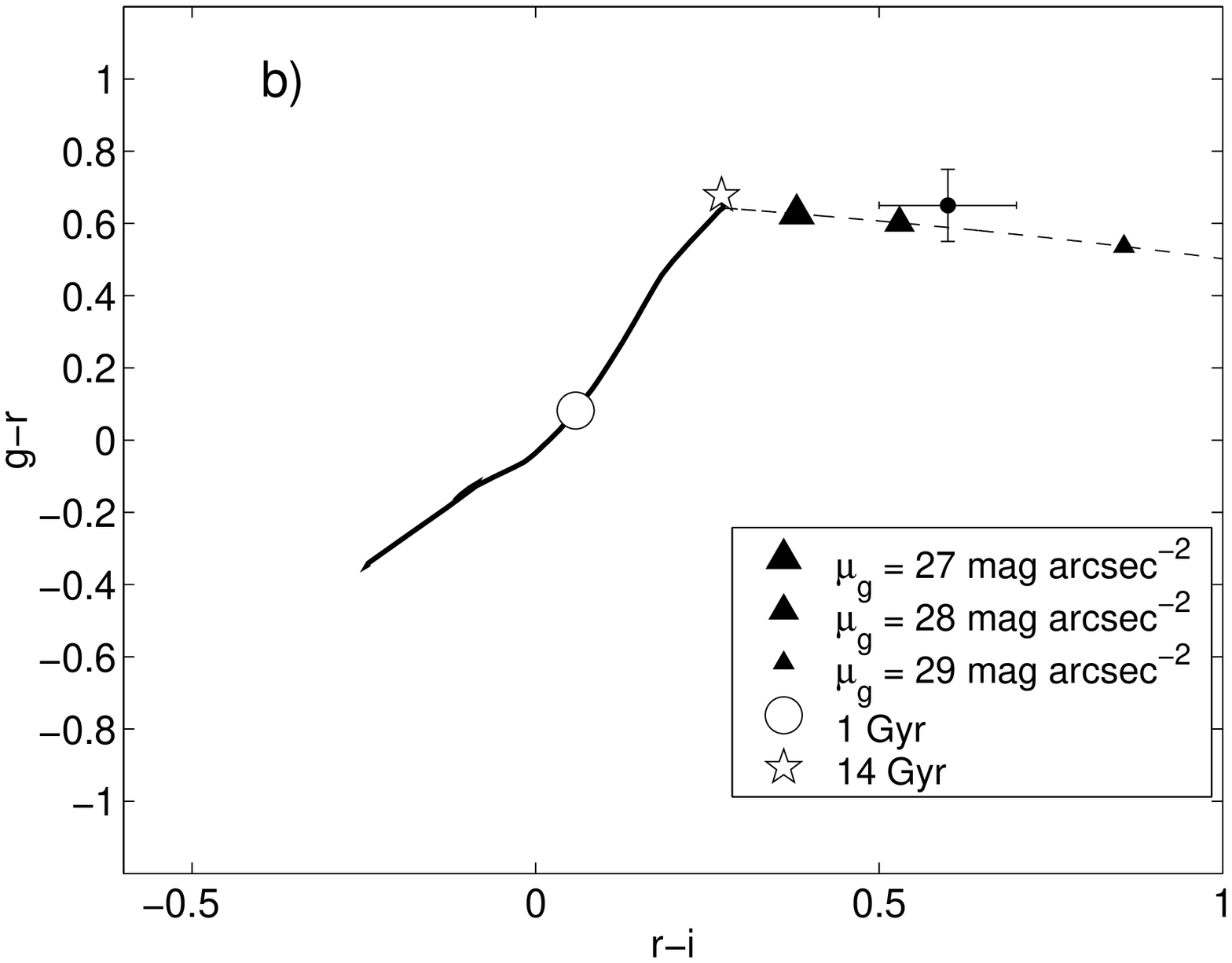}
\caption{{\bf a)} The effect of EBL extinction in a diagram of $r-i$ vs. $g-r$. The thick solid line indicates the evolution of a metal-poor ($Z=0.001$) stellar population (thick solid) with a Salpeter initial mass function and an exponentially decaying star formation rate ($\mathrm{SFR}\propto\exp{-t/\tau}$ with $\tau=1$ Gyr), based on \citet{Marigo et al.} isochrones. White markers represent ages of 1 Gyr (circle) and 14 Gyr (star). The cross represents the 1$\sigma$ error bars on the colours of the halo detected by \citet{Zibetti et al. a} around stacked edge-on discs in the SDSS (at $\mu_g\approx 28$ mag arcsec$^{-2}$). Dashed lines indicate how a spatially constant EBL extinction of $A(g)=0.1$ mag would shift the observed colours away from the model predictions under the assumption of a Milky Way extinction curve. The differently-sized triangles along these tracks correspond to surface brightness levels of $\mu_g=27$, 28, 29 and 30 mag arcsec$^{-2}$ (largest to smallest triangles). EBL extinction can indeed produce colours close to those observed, but only at relatively faint surface brightness levels ($\mu_g\approx 30$ mag arcsec$^{-2}$). {\bf b)} Same as a) but for $A_g=0.2$ mag of EBL extinction and a diffuse EBL boosted to $\mu_r=24.9$ mag arcsec$^{-2}$ and $\mu_i=24.6$ mag arcsec$^{-2}$ (at the bright end of what is allowed by the error bars on the total EBL in Table 1). In this case, the observed halo can be approximately reproduced at the surface brightness level $\mu_g\approx 28$ mag arcsec$^{-2}$ at which they were reported by \citet{Zibetti et al. a}. The age of the halo population affected by EBL extinction is here set to 9 Gyr.}
\label{fig8}
\end{figure*}

Fig.~\ref{fig8}a indicates how the $r-i$ vs. $g-r$ colours of an old, metal-poor ($Z=0.001$) stellar halo population modelled using \citet{Marigo et al.} isochrones with a Salpeter initial mass function would be affected in the presence of $A(g)=0.1$ mag extinction and a Milky Way extinction curve. The halo colours (cross indicating 1$\sigma$ error bars) reported by \citet{Zibetti et al. a} are far redder in $r-i$ than the stellar population model at all ages, but EBL extinction effects may introduce radial colour gradients (dashed lines) which substantially improve the agreement between the model and observations. 

Fig.~\ref{fig8}a reveals a number of interesting features of the EBL extinction effect. Whereas normal dust reddening of a stellar population would shift the colours along a straight vector in a colour-colour diagram like this, EBL extinction shifts the colours along curved trajectories. Moreover, the length and direction of the colour shift depends on the intrinsic colours of the affected stellar population. This is illustrated by the three different offset vectors, originating from slightly different positions along the evolutionary track of the model stellar population. The reason for this behaviour is the dependence of the colour shift on the ratio between the galaxy and EBL fluxes in equation (\ref{coloureq}) -- if the colours of the galaxy change, so will the colour shift. Similar effects can also be produced by altering the colours of the EBL itself. The triangles along each offset track (dashed line) indicate the $g$-band surface brightness at which a particular observed colour has been reached. 

While EBL extinction clearly can reproduce the observed halo colours, this happens at a surface brightness level ($\mu_g\approx 30$ mag arcsec$^{-2}$) significantly fainter than that at which the plotted colours have been measured ($\mu_g\approx 28$ mag arcsec$^{-2}$). To reconcile the reported halo colours with the predictions of EBL extinction effects, one would have to boost the amount of extinction in the halo or alter the spectral energy distribution of the EBL. Adopting a gray extinction curve also alters the colour shift, but typically not in the desired direction. Our tests indicate that, to reproduce the halo colours at $\mu_g\approx 28$ mag arcsec$^{-2}$ while keeping the surface brightness of the EBL intact would require an EBL extinction of $A(g)=1.0$ mag in the halo. Given that the \citet{Zibetti et al. a} halo colours are derived in a region located at a substantial distance away from the edge-on disc (a projected distance of twice the disc scale length), this amount of extinction seems unrealistically high. 

However, the halo colours can in principle be reproduced with an EBL extinction of $A(g)=0.2$ mag, if the total surface brightness of the EBL in the $r$ and $i$ bands is assumed to lie at the bright end of what is allowed by the error bars (this boosts the total EBL to $\mu_r=24.8$ mag arcsec$^{-2}$ and $\mu_i=24.5$ mag arcsec$^{-2}$ and the diffuse EBL to $\mu_r=24.9$ mag arcsec$^{-2}$ and $\mu_i=24.6$ mag arcsec$^{-2}$). This is demonstrated in Fig.~\ref{fig8}b. Assuming the dust to be smoothly and uniformly distributed, and the halo stars to be isotropically distributed around the halo centre, one would expect the extinction optical depth experienced by the halo stars to be approximately half of that suffered by the EBL, meaning that $A(g)=0.2$ mag for the EBL corresponds to $A(g)=0.1$ for the halo stars, or a normal dust reddening of only $E(g-r)\approx 0.03$ mag and $E(r-i)\approx 0.02$ mag (too small to be plotted in Fig.~\ref{fig8}).

As far as we know, there are no observational constraints that reliably rule out this amount of dust extinction in the halo. Dust is well known to exist a few kpc above the plane of edge-on discs \citep[e.g.][]{Alton et al.}, but constraints on the the opacity as a function of extraplanar distance are scarce. By studying the reddening of background galaxies, \citet{Zaritsky} claimed a tentative detection of $A(B)\approx 0.01$ mag of extinction at 60 kpc into the halo of two nearby disc galaxies. More recently, \citet{Menard et al.} studied the reddening of quasars projected close to galaxies in the SDSS and derived an opacity profile for galactic halos from projected distances of 20 kpc out to several Mpc. At the innermost data point ($\approx 20$ kpc), the measured opacity was no more than $A(V)\approx 0.03$ mag. Considering that the halo colours of \citet{Zibetti et al. a} were derived  $\approx 8$ kpc into the halo (based on the median scale length for galaxies in their sample), the extinction at that projected distance may of course be considerably higher. One would, however, naively expect the opacity of the halo to be lower than in a disc, and given the model described in section 4, the typical disc opacity at this surface brightness level would be $A(g)<0.1$ mag, i.e somewhat lower than required to attribute the red excess of the halo solely to EBL extinction. 

Our treatment admittedly neglects the effects of instrumental scattering on the observed surface brightness profile of the halo. Due to this effect, light from the central disc will be scattered outward and contaminate the halo region. The wavelength dependence of the instrumental point spread function (PSF) may by itself introduce spurious colour gradients and \cite{de Jong} argues that the halo signal reported by \citet{Zibetti et al. a} may be severely affected by this. If this is the case, then the light observed at any given radius in the halo will consist of at least two components: scattered light from the inner regions of the galaxy, where the dust content is likely to be higher but where the high galaxy-to-EBL flux ratio ensures essentially no EBL extinction effects, and very faint (fainter than observed) halo light which has suffered more EBL extinction effects than suggested by the observed surface brightness. Without a detailed model of the effective PSF, it becomes very difficult to reliably predict how this will alter the effects of EBL extinction.   

One could argue that, if EBL extinction is the cause of the anomalously red halos of edge-on disc galaxies, then strange colours should also be seen at a similar surface brightness level in the outer parts of discs themselves. In fact, one would expect the dust extinction to be higher in the disc than in the halo, thereby making the spurious colour gradients along the disc even more pronounced. While it would be worthwhile to search for this effect, one should bear in mind that the disc is likely to have intrinsic colours that are different from the halo (thereby altering the colour offset vector, see Fig.~\ref{fig8}a). The structure of the interstellar medium may also be different in the disc and halo, with implications for the grayness of the effective attenuation.

As seen in Figs.~\ref{fig8}a and b, EBL extinction predicts that the colours should become more and more extreme as one approaches fainter and fainter isophotes. The detection of a radially steady halo colour would therefore speak against the idea of EBL extinction as the main cause of the red excess, unless one assumes a radial evolution in the dust reddening properties which happens to cancel the colour gradient. Of course, the {\it intrinsic} halo colour may not be radially constant, since stellar halos are likely to contain age and metallicity gradients at some level.

\section{Impact on studies of the hosts of low-redshift starburst galaxies}
\begin{figure*}
\includegraphics[scale=0.45]{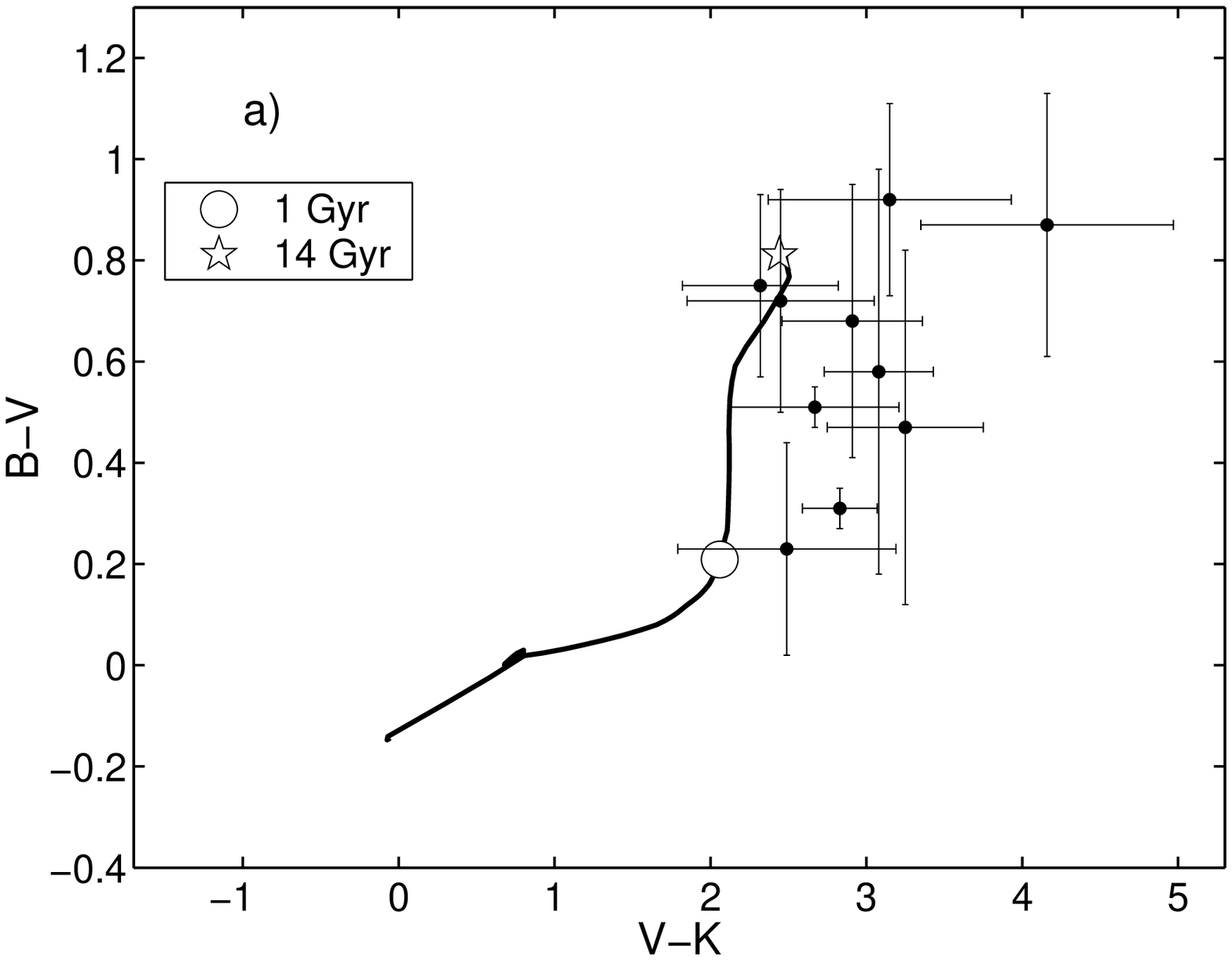}
\includegraphics[scale=0.45]{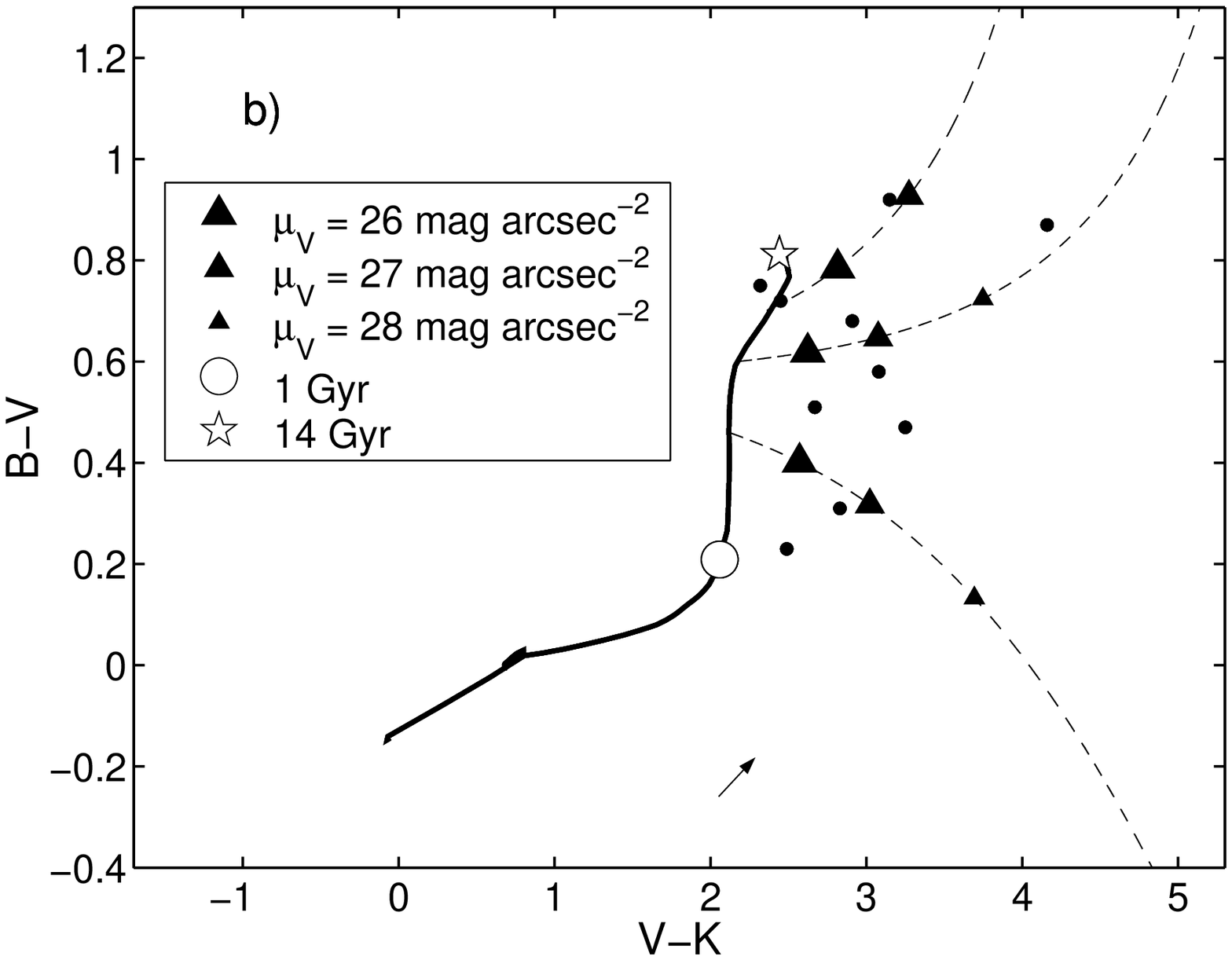}
\caption{{\bf a)} Observed $V-K$ and $B-V$ colours of the faint outskirts of BCGs (crosses indicating $1\sigma$ error bars), compared to model predictions for a stellar population with metallicity $Z=0.001$ and a Salpeter initial mass function, based on \citet{Marigo et al.} isochrones. White markers represent ages of 1 Gyr (circle) and 14 Gyr (star). The data points are clearly displaced to the right of the model tracks, indicating a $K$-band excess in the spectra of these objects. An exponentially declining star formation rate (SFR$(t)\propto \exp{-t/\tau}$) with $\tau=1$ Gyr has been assumed, but using a different star formation history does not remove the apparent red excess. {\bf b)} Same, but with error bars on the observed colours omitted to avoid cluttering. The dashed lines indicate how EBL extinction would shift the colours away from the model track in the case of $A(V)=0.5$ mag and a Milky Way extinction curve. The differently-sized triangles along these tracks correspond to surface brightness levels of $\mu_V=26$, 27 and 28 mag arcsec$^{-2}$ (largest to smallest triangles). For the chocen opacity,  EBL extinction can in principle explain the observed colours at $\mu_V\approx 26$--27. The arrow in the lower part of the figure indicates the normal dust reddening vector resulting from this scenario.}
\label{fig9}
\end{figure*}
Blue compact galaxies (BCGs) are metal-poor, low-mass systems currently undergoing intense star formation activity \citep[e.g][]{Kunth & Östlin}. While the progenitors of these objects remain unidentified, deep surface photometry of the regions outside their bright star-forming centres have provided important clues on the properties of the underlying stellar population \citep[e.g.][]{Noeske et al.,Caon et al.}. Combined optical and near-IR surface photometry \citep{Bergvall & Östlin,Bergvall et al.} have for some of these systems indicated colours that become progressively redder as one traces the surface brightness profile outwards, eventually reaching values that are difficult to reconcile with the type of stellar population one expects BCG host galaxies to have. This may be a signature of a host galaxy with a metallicity higher than the central starburst \citep{Bergvall & Östlin,Zackrisson et al. b}, or possibly a halo with a bottom-heavy initial mass function \citep{Zackrisson et al. a} similar to that advocated to explain the colours reported for the halos of edge-on disc galaxies. However, since these extreme colours turn up at surface brightnesses ($\mu_V\approx 26$--27 mag arcsec$^{-2}$; $\mu_K\approx 22$--23 mag arcsec$^{-2}$) comparable to or below the EBL level, we here investigate whether EBL extincition could possibly be responsible.

In Fig.~\ref{fig9}a, we compare the $V-K$ vs. $B-V$ colours of the faint outskirts of BCGs \citep[as presented by][]{Bergvall et al.} to the predicted evolution of a $Z=0.001$ (close to the typical metallicity of $\approx 10\%$ solar measured from emission-line ratios towards the central regions of BCGs) stellar population, based on \citet{Marigo et al.} isochrones. A Salpeter initial mass function and an exponentially decaying star formation rate ($\mathrm{SFR}(t)\propto \exp(-t/\tau)$ with $\tau=1$ Gyr) has been assumed. Many of the data points are 0.5--1.0 mag redder in $V-K$ than the model tracks, indicating either a higher host metallicity, non-negligible dust reddening or an extreme initial mass function \citep[for a recent discussion, see][]{Zackrisson et al. b}. In Fig.~\ref{fig9}b, we demonstrate how an EBL extinction of $A(V)=0.5$ mag would shift the colours away from the $Z=0.001$ track, under the assumption of a Milky Way extinction law. In this diagram, the error bars on the observed colours have been omitted to avoid cluttering. While the direction and magnitude of the shift is sensitive to the intrinsic colours of the stellar population (analogously to the situation in  Fig.~\ref{fig8}a), most of the observed colours can in principle be explained as due to EBL extinction effects. The one exception is the object at $V-K\approx 4$ (Haro 11), but new data on this galaxy indicates a problem with the original measurement, and that the redness of its outskirts may have been overestimated \citep{Micheva et al.}. 

Assuming that the EBL experiences an opacity twice that of the stars in the host, an EBL extinction of $A(V)=0.5$ mag would correspond to normal dust reddening values of $E(B-V)\approx 0.08$ mag and $E(V-K)\approx 0.22$ mag (this reddening vector has been indicated with an arrow in the lower part of Fig~\ref{fig9}b). This is within the limit of $E(B-V)< 0.1$ mag set by \citet{Bergvall & Östlin} through observations of the $\mathrm{H}\alpha/\mathrm{H}\beta$ emission-line ratio towards the central starbursts of some of these objects. While there are no direct constraints on the dust extinction in BCGs at large projected distances from their centres, one would expect the extinction to be lower in the outskirts of these objects. At a surface brightness similar to that of the region where the colours are measured ($\mu_V\approx 27$ mag arcsec$^{-2}$), the disc opacity model adopted in section 4 would for instance give an EBL extinction of only $A(V)=0.1$ mag. On the other hand, these are starburst galaxies, and the dust processing cycle may not behave the same way as in quiescent discs. Advocating $E(B-V)\approx 0.08$ mag for the BCG host galaxy would certainly not seem any more controversial than assuming a metallicity higher than observed, or a bottom-heavy initial mass function. Hence, we conclude that EBL extinction remains a viable explanation for the red excess seen in the outskirts of BCGs. 

\section{Impact on studies of intragroup and intracluster light}
Optical surface photometry of galaxy groups \citep[e.g.][]{White et al.,Da Rocha & Mendes de Oliveira,Da Rocha et al.} and clusters \citep*[e.g.][]{Mihos et al.,Gonzales et al.,Zibetti et al. b,Seigar et al.} have revealed a faint intergalactic flux component, stemming from stars liberated from tidally shredded galaxies in these dense environments \citep[][]{Purcell et al.}. Since current measurements reach as deep as $\mu_V\approx 28.5$ for individual systems \citep{Mihos et al.} and $\mu_g\approx 32$ mag arcsec$^{-2}$ for stacked clusters \citep{Zibetti et al. b}, EBL extinction effects may well be an issue. As demonstrated in previous sections, the importance of such effects strongly depend on the amount of dust present at these faint surface brightness levels. Significant dust reddening of background galaxies have been reported for individual systems -- most notably the M81 group \citep[corresponding to $A(V)\approx 0.4$ mag if a Milky way extinction curve is assumed;][]{Xilouris et al.}, where the starburst galaxy M82 is ejecting dust into the intragroup medium. However, several investigations aiming to measure the reddening of galaxies and quasars behind large samples of clusters have reported no or very little average reddening \citep*[$A(V)\leq 0.06$ mag, assuming Milky Way extinction;][]{Chelouche et al.,Muller et al.,Bovy et al.}. A highly irregular dust distribution may in principle give rise to sizable extinction even if there is little reddening (as appears to be the case in disc galaxies, see \citealt{Keel & White, Holwerda et al. a}), but without detailed constraints on the grayness of the attenuation, we are forced to assume a Milky Way extinction law in the following.

\begin{figure}
\includegraphics[width=84mm]{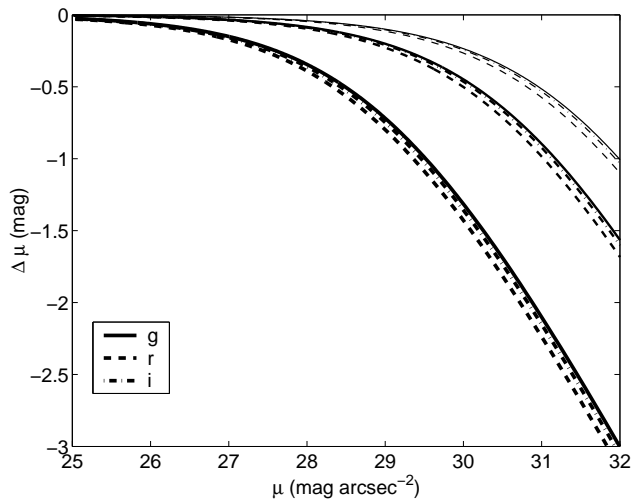}
\caption{The offset $\Delta\mu$ between the intrinsic and observed surface brightness, as a function of observed surface brightness $\mu$, in filters $g$ (solid lines), $r$ (dashed lines) and $i$ (dash-dotted lines) predicted by EBL extinction effects over the range of surface brightnesses relevant for current measurements of intragroup and intracluster light. Line thicknesses indicate EBL extinction values of $A(V)=0.05$ mag (thick lines), 0.01 mag (medium lines) and 0.005 mag (thin lines), assuming a Milky Way extinction curve.}
\label{fig10}
\end{figure}

In Fig.~\ref{fig10}, we show the offset $\Delta\mu$ between the intrinsic and observed surface brightness, as a function of observed surface brightness $\mu$ (i.e. the surface brightness inferred when EBL extinction is neglected), in filters $g$ (solid lines), $r$ (dashed lines) and $i$ (dash-dotted lines) predicted for an EBL extinction of $A(V)=0.05$ mag (thick lines), 0.01 mag (medium lines) and 0.005 mag (thin lines). The difference between Fig.~\ref{fig10} and Fig.~\ref{fig4} is that Fig.~\ref{fig10} uses SDSS filters and dust opacities more relevant for the intracluster medium.

In the case of $A(V)=0.05$ mag (thick lines), the observed $g$-band profile is $\approx 1.3$ mag fainter than the intrinsic one at an observed $\mu_g\approx 30$ mag arcsec$^{-2}$ \citep[approximately corresponding to the faint limit of the total observed light profile derived from stacked cluster images by][]{Zibetti et al. b}. Even at $\mu_g\approx 28$ mag arcsec$^{-2}$ (approximately the current limit of detections of intracluster light in individual clusters), the offset is $\approx 0.3$ mag. The effect becomes smaller for smaller $A(V)$, and  becomes irrelevant ($<0.05$ mag) for $A(V)=0.005$ mag at $\mu_g\approx 28$ mag arcsec$^{-2}$. 

\section{A new route to constraining the surface brightness of the extragalactic background}
While EBL extinction appears to set a limit to how deep surface photometry can be pushed without introducing serious systematic errors, this effect need not only be a source of nuisance. As described in the introduction, there are two complementing methods for studying regions of low surface brightness: surface photometry and direct star counts. The radius out to which star counts can be trusted is primarily set by the level of contamination from foreground stars and background galaxies, rather than the brightness of the night sky. Hence, EBL extinction effects should not be an issue for surface brightness profiles based on this technique. This allows for an independent way of measuring (or setting an upper limit on) the surface brightness of the EBL, which could help resolve the controversy regarding current measurements (as described in section 3.).

If surface photometry reveals a significant drop in surface brightness (as in Fig.~\ref{fig3}a, Fig.~\ref{fig5}a, Fig.~\ref{fig6} and Fig.~\ref{fig7}), whereas star counts indicate no such feature, this would be a tell-tale sign of EBL extinction effects. By looking for such discrepancies in systems with known opacity profiles, the surface brightness of the EBL in the filter used can directly be estimated from equation (\ref{photeq}). If, on the other hand, no such discrepancies between the two methods are detected, this can be used to set an upper limit on the EBL.

The best targets for an endeavour of this type would seem to be disc galaxies, both because these have relatively high opacities \citep[e.g.][]{Holwerda et al. a} and also since average opacity profiles have already been derived \citep[e.g][]{Holwerda et al. b}. Under the assumption of the disc galaxy model used in Fig.~\ref{fig5}, the absence of any discrepancy between the surface brightness profiles derived from surface photometry and star counts (at the $\geq 0.1$ mag level) out to $\mu_V\approx 28$ mag arcsec$^{-2}$ would indicate that the diffuse $V$-band surface brightness of the diffuse EBL is a factor of 5 lower (i.e. 1.75 mag fainter) than indicated by the direct measurements (Table 1). By combining surface photometry, star counts and opacity measurements (through counts of background galaxies), a test of this type can in principle be implemented. The weakest link is likely to be the opacity measurements, since current techniques tend to yield rather large error bars on the extinction at faint isophotes \citep{Holwerda et al. a}. Combined surface photometry and star counts are already available for a small number of disc galaxies, for instance NGC 4244 \citep{Fry et al.,de Jong et al.}, but the edge-on orientation of this particular object makes it far from optimal. An alternative strategy could be to analyze a sizable number of discs without individual opacity measurements, since the extinction profile of \citet{Holwerda et al. b} is likely to hold on average.

\section{Discussion}
\subsection{The surface brightness of the diffuse EBL}
The quantitative EBL extinction effects discussed in previous sections are based on the surface brightness of the diffuse (spatially unresolved) EBL, as estimated in section 3. Since the diffuse EBL is derived by subtracting the resolved EBL from the total, we caution that the surface brightness of this component may change dramatically, if either the total EBL turns out to be overestimated (as implied by estimates of the opacity of the Universe at gamma-ray wavelengths; e.g. \citealt*{Franceschini et al.,Mazin & Raue,Albert et al.,Razzaque et al.}) or, alternatively, if the EBL that can be attributed to resolved galaxies has been underestimated. 

As shown in section 3, the current difference between the total and diffuse EBL is only 0.1--0.5 mag. This means that the contribution from resolved sources is minor and that EBL extinction effects are fairly insensitive to the masking of background galaxies. This may no longer be the case if the total EBL level is significantly lower than implied by current direct measurements. In principle, the diffuse component may tend towards zero if future measurements are able to completely reconcile the total EBL with resolved galaxies. We point out, however, that a too rigorous masking of faint background sources may give rise to unwanted effects on its own. Due to contrast issues, background sources will be more difficult to identify when superposed on faint isophotes of the target object than when located at greater projected distances in the sky. Hence, it may be impossible to reliably mask out interlopers projected against the target object as faint as those that can be masked in sky fields. The EBL level estimated after masking resolved sources well away from the target galaxy will thereby be too low, leading to an undersubtraction of sky. This effect may to some extent be canceled by extinction within the target object, which makes background objects appear dimmer, but to estimate the net result on colour maps and surface brightness profiles would require very careful modelling. 

\subsection{Spatial fluctuations in the EBL}
Significant spatial fluctuations in the EBL have been reported on scales of $\sim 1$\arcsec{ }and upwards at 1--2 $\mu$m \citep[e.g.][]{Kashlinsky et al.,Odenwald et al.,Matsumoto et al.,Thompson et al.}. This means that the effects of EBL extinction are likely to depend on the angular size of the objects studied, since the EBL contribution to the overall sky flux may vary from target to target or across the face of a single object. This effect is difficult to quantify without knowledge of the detailed angular fluctuation spectrum as a function of wavelength, but it should be kept in mind that this may introduce significant uncertainties in the predictions for EBL extinction effects for individual objects. 

\subsection{Impact of EBL extinction on studies of high-redshift galaxies}
As discussed in section 3, the discrepancy between direct EBL measurements and the estimated contribution from resolved galaxies indicates that most of the EBL is diffuse (or currently unresolved). Current models moreover have great difficulties in reproducing the observed properties of the optical/near-IR EBL without resorting to exotic scenarios \citep[e.g.][]{Matsumoto et al.,Fardal et al.}. Hence, the origin of the EBL and its redshift evolution remains poorly constrained. This makes it very difficult to assess the effect of EBL extinction on surface photometry of objects beyond the local Universe, since part of the EBL may then constitute a foreground, rather than a background, and would not give rise to effects such as those discussed in previous sections.

\subsection{Impact of EBL extinction on studies of point-like objects}
So far, we have discussed the impact of EBL extinction on surface photometry of extended objects, but faint point sources could in principle also be subject to similar effects. If the part of the EBL that current telescopes are unable to resolve is due to faint, extended wings of distinct astronomical objects or to some unknown mechanism in the intergalactic medium (i.e. a truly diffuse EBL), point sources with dust structure on the scale of the seeing disc could be affected in a way similar to that of extended objects. Once the surface brightness across the seeing disc becomes similar or lower than that of the EBL, an overdensity of dust correlated with this seeing disc (dust either in the direct vicinity of the point source or in the foreground) would make estimates of the sky flux (based on measurements {\it outside} the seeing disc) too high. This results in an oversubtraction of sky, in direct analogy with the effect discussed in previous sections. At such small angular scales, the issue of spatial variations in the EBL of course becomes severe, making it very difficult to assess the likely impact of EBL extinction on realistic sources.  

\subsection{Impact of EBL extinction on studies of Milky Way objects}
This paper has focused on the impact of EBL extinction on surface photometry of extragalactic objects, but this mechanism should also affect surface photometry measurements on more nearby objects, such as star clusters and nebulae within the Milky Way. The resulting oversubtraction of sky may in this case be even more severe, since the EBL does not necessarily represent the only sky component coming from behind the target objects -- faint Milky Way stars or diffuse light from the background interstellar medium may also contribute.

\section{Summary}
Our conclusions can be summarized as follows:
\begin{enumerate}
\item We have pointed out a previously unrecognized problem with surface photometry at the faint limit, related to how the sky flux is typically estimated and subtracted from astronomical images. Since the EBL part of the sky flux is likely to originate from behind extended objects in the low-redshift Universe, the overall sky flux across such targets may be diminshed due to extinction by dust. This causes a systematic oversubtraction of sky, which will affect surface brightnesses similar to or fainter than that of the diffuse (unresolved) component of the EBL. Without detailed knowledge of the opacities of the objects studied, this effect sets a limit to how deep surface photometry can be pushed without introduction of serious systematic errors.
\item We find that EBL extinction effects can mimic the truncation of disc galaxies, but is unlikely to be the cause of such features in general, except maybe for the dustiest discs.
\item EBL extinction may provide a partial explanation for the unexpectedly red colours reported for the halos of disc galaxies and hosts of local starburst galaxies. To attribute the colour anomalies solely to this effect would, however, require more dust than naively expected in the faint outskirts of such systems.
\item Based on current constraints on the dust content of galaxy groups and clusters, EBL extinction may lead to a non-negligible underestimate of intragroup and intracluster light at the faintest surface brightness levels currently probed.
\item By combining surface photometry and direct star counts for systems with reasonably well-constrained opacity profiles, it should be possible to set an independent constraint on the EBL. This could help resolve current controversies regarding the surface brightness of the EBL.
\end{enumerate}

\section*{Acknowledgments}
EZ acknowledges research grants from the Swedish Research Council, the Royal Swedish Academy of Sciences and the Royal Physiographical Society of Lund.

\end{document}